\documentclass[english]{IEEEtran}
\usepackage[T1]{fontenc}
\usepackage[latin9]{inputenc}
\usepackage{amsthm}
\usepackage{amsmath}
\usepackage{amssymb}
\usepackage{graphicx}

\makeatletter
\theoremstyle{plain}
\newtheorem{thm}{\protect\theoremname}
\theoremstyle{definition}
\newtheorem{example}[thm]{\protect\examplename}
\theoremstyle{definition}
\newtheorem{defn}[thm]{\protect\definitionname}
\theoremstyle{plain}
\newtheorem{lem}[thm]{\protect\lemmaname}

\makeatother

\usepackage{babel}
\providecommand{\definitionname}{Definition}
\providecommand{\examplename}{Example}
\providecommand{\lemmaname}{Lemma}
\providecommand{\theoremname}{Theorem}

\begin{document}

\title{Theoretic Shaping Bounds for Single Letter Constraints and Mismatched
Decoding}

\author{Stella Achtenberg, \emph{Student Member, IEEE,} and Dan Raphaeli,
\emph{Member, IEEE}}
\maketitle
\begin{abstract}
Shaping gain is attained in schemes where a shaped subcode is chosen
from a larger codebook by a codeword selection process. This includes
the popular method of Trellis Shaping (TS), originally proposed by
Forney for average power reduction. The decoding process of such schemes
is mismatched, since it is aware of only the large codebook. This
study models such schemes by a random code construction and derives
achievable bounds on the transmission rate under matched and mismatched
decoding. For matched decoding the bound is obtained using a modified
asymptotic equipartition property (AEP) theorem derived to suit this
particular code construction. For mismatched decoding, relying on
the large codebook performance is generally wrong, since the performance
of the non-typical codewords within the large codebook may differ
substantially from the typical ones. Hence, we present two novel lower
bounds on the capacity under mismatched decoding. The first is based
upon Gallager's random exponent, whereas the second on a modified
version of the joint-typicality decoder. \end{abstract}

\section{Introduction}

The problem of efficient communication has been extensively studied
for a variety of channels and constraints. Shannon has found the explicit
channel capacity of the additive white Gaussian noise (AWGN) channel
subject to average power constraint \cite{key-12}. This capacity
is achieved by a continuous Gaussian input. In practice, however,
discrete constellations are often used for transmission. In this case,
the probability mass function which maximizes the achievable rate
subject to a given discrete constellation and average power constraint
can be obtained numerically using the modified \emph{Blahut-Arimoto}
Algorithm (BAA) \cite{key-8}.\textbf{ }The binary symmetric and non-symmetric
channels (BSC and BNSC, respectively) subject to hamming power (bounded
number of '1') constraint are also important examples, with various
applications. For BSC capacity and applications, see \cite{key-4}
and \cite{key-5}. For the capacity of the noise free aperture channel
(free space optical communication), modeled by a special case of BNSC
(Z-channel), see \cite{key-15}. All these examples share single letter
constraints and memoryless channels. Such constraints and channels
are treated in this paper.

Practical shaping methods have been devised over the years to approach
channel capacity under constraints. Convey and Slone \cite{key-2}
used nested lattice codes for constellation shaping. Their construction
was generalized by Forney \cite{key-3}, subsequently leading to the
trellis shaping technique for average power reduction \cite{key-1}.
In this method, a structured large code is formed by nesting the shaping
code with the channel code. Then, given the coded bits, the free shaping
bits creates a set of sequences, i.e., a coset of the large code,
from which one sequence is chosen such that the transmitted codeword
has minimal power in the coset. This operation is very similar to
choosing a codeword which satisfies a constraint as will be shown
in the sequel. The chosen codewords from each coset form a shaped
subcode, which achieves shaping gain. 

Another example of such shaped code construction is the LDPC-LDGM
scheme \cite{key-6}, originally designed for dirty paper coding.
This scheme creates a shaping code for each transmitted message using
a nested LDPC-LDGM structure, which is designed to be both good source
code and good channel code, thus shaping is possible. The shaping
codes serve as cosets. Once again, the minimum energy codeword is
chosen from each coset, forming the shaped subcode. The large code,
as before, in the union of all the cosets representing each message. 

All these schemes share a mismatched decoding process. Their decoder
is only aware of the large codebook, unaware of the constraints and
performs optimal decoding as if all codewords might have been transmitted.
This decoder is suboptimal, but occasionally more practical since
the adaptation to the shaped subcode might be extremely complex, and
the decoder for the large code is readily available. We refer to it
as the \emph{mismatched decoder}. The optimal decoder, which is adapted
to the shaped subcode, i.e., familiar with the constraint and is able
to repeat the selection process at the receiver, is referred as the
\emph{matched decoder}. 

Although these works are practical and demonstrate significant shaping
gains, the analysis of their performance limit is partial. In \cite{key-1},
Forney uses {}``random shaping'' arguments to suggest why the trellis
shaping approaches the ultimate shaping gain so rapidly. These arguments,
however, are based upon geometrical volumes analysis, which is more
suitable for large constellations and high SNR. The achievable rate
using the matched decoder and the shaped code distribution are not
addressed. For the mismatched decoding, Forney and others rely on
the large codebook performance. On first look, mismatch performance
analysis seems as a trivial exercise. The transmitted codeword is
taken from the large codebook, and therefore the large codebook performance
can be used. This analysis is referred to as a \emph{naive approach}.
The large code performance analysis, however, assumes the transmission
of the typical codewords, while in the shaping system we transmit
(almost) only non-typical codewords. These codewords may have worse
distance spectrum towards their neighbors and error free transmission
is not guaranteed at the same noise level. Thus, a more precise analysis
is necessary.

In this paper, we construct random code ensembles, which satisfy constraints,
via codeword selection process. The large codebook and the cosets
within are generated at random from an i.i.d. distribution. In this
case the {}``cosets'' are random, referred from this point on as
random sets. Our code construction provides a framework for performance
analysis of practical shaping schemes, since both good channels codes
and good shaping codes behave approximately as random codes. In this
paper, we find the minimal size of a random set needed to ensure that
given a set of constraints, at least one codeword satisfies the constraints
in each random set with probability tending towards 1. Furthermore,
we obtain bounds on the achievable rates of the matched and the mismatched
decoders subject to single letter constraints.

\begin{figure}
\begin{centering}
\includegraphics[width=8cm,height=8cm,keepaspectratio]{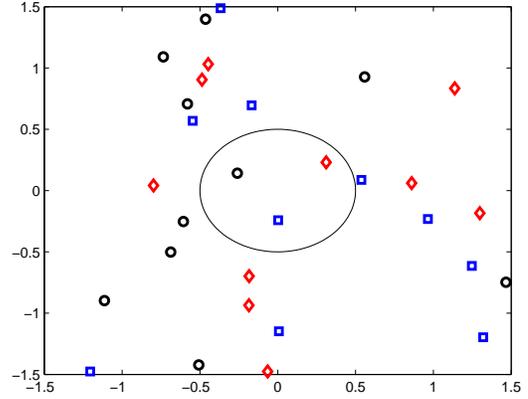}
\par\end{centering}

\caption{\label{fig:Random-subcode-selection}Random subcode construction process.}
\end{figure}

More formally, let $N$ be the transmission block length and $R_{q}$
be the transmission rate. Let $\mathcal{C}$ be a random codebook
with rate $R$, referred to as the \emph{large codebook}. Let $R_{s}$
be a design parameter, called \emph{shaping rate}, such that $R=R_{q}+R_{s}$.
The large codebook is the union of $2^{NR_{q}}$ random sets, with
$2^{NR_{s}}$ codewords in each set, where each codeword is generated
according to an i.i.d. probability mass function $p\left(x\right)$.
A message $\boldsymbol{m}$ to be transmitted chooses the random set.
Out of this random set a single codeword $\boldsymbol{X}=\left(X_{0},\ldots,X_{N-1}\right)\in\mathcal{C}$
is selected such that it satisfies a set of single letter constraints.
In this case the constraints are codeword related and can be formulated
as
\begin{equation}
\frac{1}{N}\sum_{i=0}^{N-1}\varphi_{l}\left(X_{i}\right)\leq\beta_{l},
\end{equation}
where $l=0..L-1$, $L$ is the number of constraints, $\beta_{l}$
are constraints and $\varphi_{l}\left(\cdot\right)$ are bounded constraints
functions. The union of the selected codewords is the shaped code
$\mathcal{C}_{s}$. Since each codeword satisfies the constraints,
the codebook satisfies the constraints in average. The shaping rate
$R_{s}$, which determines the random set size, should be chosen such
that a random set includes at least one suitable codeword with probability
approaching 1. A simple illustration of the selection process can
be seen in Fig. \ref{fig:Random-subcode-selection}, where the large
codebook is the union of the three random sets, designated by the
different colors and shapes. The constraint is plotted as the circle
and one codeword is chosen from each random set such that it is inside
the circle, i.e., satisfies the constraint. These three codewords
construct the shaped code. Another simple example of a constraint
and a selection process is as follows. 
\begin{example}
Let the {}``constellation'' be binary and the constraint be on the
average number of '1' within the codebook, say limited to $\frac{1}{3}$.
To construct the shaped codebook, let the large codebook be binary
and i.i.d. according to $X_{i}\sim Bernoulli\left(\frac{1}{2}\right)$.
We choose codewords that satisfy this constraint. Thus, each codeword
in the random set undergoes the following test 
\begin{align}
 & \frac{1}{N}\sum_{i=0}^{N-1}X_{i}\leq\frac{1}{3},\label{eq:hamming const 1/3}
\end{align}
using $\varphi\left(x\right)=x$. If the average number of '1' in
the codeword is limited to $\frac{1}{3}$, then it is chosen for the
shaped codebook. For example, the codeword $\left(1,0,1,0,1,0,1,0,1,0,1,0\right)$
is typical in the large code, since the number of zeros and ones is
equal. However, it does not satisfy the constraint. On the other hand,
the non-typical codeword $\left(1,0,0,1,0,0,1,0,0,1,0,0\right)$ satisfies
the constraint. 
\end{example}
Thus, the questions in hand are how to choose the shaping rate $R_{s}$
to find such non-typical codewords, and what are the achievable rates
using this code construction under the matched and the mismatched
decoders.
 The Large Deviation Theory, which includes the Sanov theorem and
the conditional limit theorem \cite{key-10}, addresses the probability
of very non-likely events, therefore provides some of the answers.
For example, the probability to generate a codeword which satisfies
the hamming constraint in (\ref{eq:hamming const 1/3}) from a uniform
distribution. This probability dictates the minimum size of each random
set, and obviously the minimal shaping rate $R_{s}$ required to ensure
that each random set includes at least one such codeword. Furthermore,
the conditional limit theorem shows that the marginals of the $N$-dimensional
distribution of the selected codewords tend toward the \emph{conditional
limit distribution}, which is an i.i.d. distribution $q^{\star}\left(x\right)$.
It is the closest distribution to $p\left(x\right)$ (in the sense
of divergence) among all the distributions which satisfy the constraints.

Further, we derive achievable bounds on the transmission rate under
matched and mismatched decoder. Starting with the matched decoder,
one of the methods to prove achievable rate for a random ensemble
of codes, which was generated according to an i.i.d. probability mass
function, is the asymptotic equipartition property (AEP) theorem \cite{key-10}.
Since the shaping method involves codeword selection and not random
generation from a target distribution, we derive a modified AEP theorem.
We bound the probability to generate a codeword via this codebook
construction using the conditional limit distribution. The requirement
that each codeword satisfies the constraints is stronger than the
requirement that the whole codebook satisfies the constraints in average;
since a small amount of codewords which do not satisfy the constraints
do not effect the average. Thus, the codebooks that are shaped using
this methods are a subset of the codebooks which satisfy the constraints
in average. Nevertheless under matched decoding, this code construction
may achieve the same rate as a code which was generated i.i.d. according
to $q^{\star}\left(x\right)$, i.e., the mutual information $I_{q^{\star}}\left(X;Y\right)$.
This property is shown using the modified AEP. 

Note that the rate achieved by the matched decoder is usually not
the channel capacity since $q^{\star}\left(x\right)$ is generally
not the distribution that maximizes the mutual information. For example,
it can be easily shown that using an i.i.d. uniform distribution for
the large code, the conditional limit distribution $q^{\star}\left(x\right)$
maximizes the entropy of the input and not the mutual information.
Hence, $q^{\star}\left(x\right)$ achieves only the lower bound on
the capacity subject to the constraints. Although it is only a lower
bound, usually it is quite tight. This conjecture is based of the
tightness of this bound for the single letter average power constraint
over the AWGN channel. This can be verified using the modified BAA
\cite{key-8}, since simulations show that the loss in rate due to
the suboptimal distribution is insignificant in all SNR regions.

Next, we derive two novel lower bounds on the capacity under mismatched
decoding. The first is based upon Gallager's random exponent.\textbf{
}The key here is to perform maximum likelihood decoding using the
entire codebook, while taking into account that only a subcode was
transmitted and the true distribution of the transmitted codewords.
Based on this analysis, we redefine the average error probability
and present a new error exponent. Using the new error exponent, we
show achievable rate for the mismatched decoder. Our second bound
is based upon a modified version of the joint-typicality decoder.
This original decoder is not suitable for non-typical codewords, since
it declares an error when the input or the output is not typical.
Thus, we define a Modified-Joint-Typicality (MJT) decoder, which decodes
solely based upon joint typicality between the input and the output
with respect to the large codebook distribution. It might seem surprising
that decoding according to a wrong distribution works and indeed this
bound is correct in special cases only. One of them is uniform distribution
of the large codebook and additive channels. Therefore, the AWGN channel
can be treated using this approach and MJT bound can be obtained. 

We show result for several examples: PAM with power constraints over
AWGN, BSC and BNSC with hamming constraints. Our bounds show that
for PAM over the AWGN channel there is no significant loss due to
mismatch decoding. Thus, shaping gain can be attained even under mismatched
decoding. This conclusion actually justifies the works of Forney and
others (unknowingly), who used mismatched decoding. The BSC channel
is an example where the naive approach works and coincides with our
bounds. It is known that the BSC capacity is attained for uniform
input probability mass function, and that linear codes achieve the
BSC capacity. Such codes have the same distance spectrum for all the
codewords and therefore the same error probability. In particular,
the error probability of the non-typical codewords equals to the error
probability of the typical. In this special case, the large codebook
performance is the same as the shaped code. This example provides
some insight to why the naive approach is correct for BSC. 

The paper is organized as follows. Section \ref{sec:Preliminaries}
presents some of the preliminary background in Large Deviations and
Method of Types. Section \ref{sec:Subcode-construction} presents
our subcode construction method and the summary of the main results
of this paper. Including the minimal shaping rate required for this
code construction is presented in Theorem \ref{thm:Number of NT codewords}.
It also addresses decoder types and examples of related practical
shaping schemes. Section \ref{sec:Matched-decoder} analyzes the matched
decoder. Its significant contribution is the modified AEP theorem,
which is necessary to establish the achievable rate under matched
decoding. Section \ref{sec:Mismatched-Decoding} analyzes the mismatched
decoder and presents two achievable bounds, the MJT bound and the
Gallager bound. Section \ref{sec:Special-cases} is dedicated to examples
of channels and constraints, such as BSC and AWGN subject to average
power constraints.

\section{\label{sec:Preliminaries}Preliminaries}

Let capital letters denote random variables and vectors. Let small
and calligraphic letters denote their realizations and support, respectively.
Bold letters represent vectors. Let $X$ be a random variable with
probability mass function $p\left(x\right)$ and support $\mathcal{X}$.
Let $p\left(y\mid x\right)$ be a discrete memoryless channel and
$Y$ be a random variable with probability mass function $p\left(y\right)$
and support $\mathcal{Y}$, such that $X$ and $Y$ are the input
and output to the channel, respectively. The joint probability mass
function of $\left(X,Y\right)$ is $p\left(x,y\right)$. For the sake
of completeness, we repeat several basic quantities, which can be
found in \cite{key-10}. Thus, the entropy of the variable $X$ is
\begin{equation}
H_{p}\left(X\right)=-\sum_{x\in\mathcal{X}}p\left(x\right)\log p\left(x\right),\label{eq:entropy}
\end{equation}
the conditional entropy of $Y$ given $X$ is
\begin{equation}
H_{p}\left(Y\mid X\right)=-\sum_{x\in\mathcal{X},y\in\mathcal{Y}}p\left(x,y\right)\log p\left(y\mid x\right)\label{eq:conditional entropy}
\end{equation}
and the mutual information of $\left(X,Y\right)$ is
\begin{equation}
I_{p}\left(X;Y\right)=H_{p}\left(Y\right)-H_{p}\left(Y\mid X\right).\label{eq:mutual information}
\end{equation}
The mutual information has also other formulations. Furthermore, let
$q\left(x\right)$ be another probability mass function on the same
support $\mathcal{X}$. The Kullback-Leibler information divergence
(also relative entropy) between $p\left(x\right)$ and $q\left(x\right)$
(not symmetric) is 
\begin{equation}
D\left(q\left(X\right)\parallel p\left(X\right)\right)=\sum_{x\in\mathcal{X}}q\left(x\right)\log\frac{q\left(x\right)}{p\left(x\right)},\label{eq:divergence}
\end{equation}
using the conventions that $0\log\frac{0}{0}=0$, $0\log\frac{0}{q}=0$
and $p\log\frac{p}{0}=\infty$.

Next we review several definitions and theorems from the AEP \cite[Chapters 3,7]{key-10}. 
\begin{defn}
\label{A_px}The typical set $A_{\epsilon,p\left(x\right)}^{(N)}$
with respect to the probability mass function $p\left(x\right)$ is
the set of sequences $\boldsymbol{x}\in\mathcal{X}^{N}$ with length
$N$ and empirical entropies $\epsilon$-close to the true entropy
$H_{p}\left(X\right)$:
\begin{eqnarray}
A_{\epsilon,p\left(x\right)}^{(N)} & = & \left\{ \boldsymbol{x}\in\mathcal{X}^{N}:\right.\nonumber \\
 &  & \left.\left|-\frac{1}{N}\log p(\boldsymbol{x})-H_{p}\left(X\right)\right|<\epsilon\right\} .\label{eq:typX}
\end{eqnarray}

\end{defn}
The AEP theorem \cite[Theorem 3.1.1]{key-10}, shows that if $\boldsymbol{X}$
are sequences of length $N$ drawn i.i.d. according to $p\left(x\right)$,
then:
\begin{eqnarray}
\Pr\left(\boldsymbol{X}\in A_{\epsilon,p\left(x\right)}^{(N)}\right) & > & 1-\epsilon,\label{eq:AEP prop X}
\end{eqnarray}
for sufficiently large $N$.
\begin{defn}
\label{A_pxy}The jointly typical set $A_{\epsilon,p\left(x,y\right)}^{(N)}$
with respect to the probability mass function $p\left(x,y\right)$
is the set of sequences $(\boldsymbol{x},\boldsymbol{y})\in\mathcal{X}^{N}\times\mathcal{Y}^{N}$
with length $N$ and empirical entropies $\epsilon$-close to the
true entropies:
\begin{eqnarray}
A_{\epsilon,p\left(x,y\right)}^{(N)} & = & \left\{ (\boldsymbol{x},\boldsymbol{y})\in\mathcal{X}^{N}\times\mathcal{Y}^{N}:\right.\nonumber \\
 &  & \;\left|-\frac{1}{N}\log p(\boldsymbol{x})-H_{p}\left(X\right)\right|<\epsilon,\nonumber \\
 &  & \;\left|-\frac{1}{N}\log p(\boldsymbol{y})-H_{p}\left(Y\right)\right|<\epsilon,\nonumber \\
 &  & \left.\left|-\frac{1}{N}\log p(\boldsymbol{x},\boldsymbol{y})-H_{p}\left(X,Y\right)\right|<\epsilon\right\} .\label{eq:typXY}
\end{eqnarray}

\end{defn}
The Joint AEP theorem \cite[Theorem 7.6.1]{key-10}, shows that if
$\left(\boldsymbol{X},\boldsymbol{Y}\right)$ are sequences of length
$N$ drawn i.i.d. according to $p\left(x,y\right)$, then:
\begin{eqnarray}
\Pr\left(\left(\boldsymbol{X},\boldsymbol{Y}\right)\in A_{\epsilon,p\left(x,y\right)}^{(N)}\right) & > & 1-\epsilon,\label{eq:AEP prop XY}
\end{eqnarray}
for sufficiently large $N$.

We also repeat some definitions and theorems from the Method of Types
given in \cite[Chapter 11]{key-10}.
\begin{defn}
The type $P_{\boldsymbol{x}}$ of a sequence $\boldsymbol{x}=\left(x_{1},\ldots,x_{N}\right)$
is the relative proportion of occurrences of each symbol of $\mathcal{X}$,
i.e., $P_{\boldsymbol{x}}\left(a\right)=N\left(a\mid\boldsymbol{x}\right)/N$
for all $a\in\mathcal{X}$, where $N\left(a\mid\boldsymbol{x}\right)$
is the number of times the symbol $a$ occurs in the sequence $\boldsymbol{x}\in\mathcal{X}^{N}$.

Let $\mathcal{P}_{N}$ be the set of types with denominator $N$ (all
possible types when the sequences have length $N$). Hence, if $P\in\mathcal{P}_{N}$
the type class $T\left(P\right)$ is the set of sequences of length
$N$ and type $P$, i.e., 
\begin{equation}
T\left(P\right)=\left\{ \boldsymbol{x}\in\mathcal{X}^{N}:\; P_{\boldsymbol{x}}=P\right\} .
\end{equation}

\end{defn}
The following properties are established in \cite[Theorems 11.1.2 - 11.1.4]{key-10}:
\begin{enumerate}
\item If $\boldsymbol{X}=\left(X_{1},\ldots,X_{N}\right)$ are drawn i.i.d.
according to $p\left(x\right)$, the probability of $\boldsymbol{x}$
depends only on its type and is given by
\begin{equation}
p^{N}\left(\boldsymbol{x}\right)=2^{-N\left(H\left(P_{\boldsymbol{x}}\right)+D(P_{\boldsymbol{x}}\parallel p\left(X\right))\right)}.\label{eq:seq. probability}
\end{equation}

\item For any type $P\in\mathcal{P}_{N}$,
\begin{equation}
\frac{2^{NH\left(P\right)}}{\left(N+1\right)^{\left|\mathcal{X}\right|}}\leq\left|T\left(P\right)\right|\leq2^{NH\left(P\right)}.
\end{equation}

\item For any type $P\in\mathcal{P}_{N}$ and probability mass function
$p\left(x\right)$, the probability of type class $T\left(P\right)$,
where the sequences are drawn according to i.i.d. $p\left(x\right)$
is given by
\begin{equation}
\frac{2^{-ND(P\parallel p\left(X\right))}}{\left(N+1\right)^{\left|\mathcal{X}\right|}}\leq p^{N}\left(T\left(P\right)\right)\leq2^{-ND(P\parallel p\left(X\right))}.\label{eq:Type probability}
\end{equation}

\end{enumerate}
The Sanov Theorem \cite[Theorem 11.4.1]{key-10} is part of the Large
Deviation Theory, which addresses the probability of very non-likely
events. The theorem is formulated as follows; let $\boldsymbol{X}=\left(X_{1},\ldots,X_{N}\right)$
be drawn i.i.d. according to $p\left(x\right)$. Let $E\subseteq\mathcal{P}$
be a set of probability mass functions. Then, the probability that
$P_{\boldsymbol{X}}$ belongs to the set $E\cap\mathcal{P}_{N}$ is
\begin{eqnarray}
p^{N}\left(E\right) & = & p^{N}\left(E\cap\mathcal{P}_{N}\right)\nonumber \\
 & \leq & \left(N+1\right)^{\left|\mathcal{X}\right|}2^{-ND\left(q^{\star}\left(X\right)\parallel p\left(X\right)\right)},\label{eq:p(N)E}
\end{eqnarray}
where 
\begin{equation}
q^{\star}\left(x\right)=\arg\min_{P\left(x\right)\in E}D\left(P\left(X\right)\parallel p\left(X\right)\right)\label{eq:q star}
\end{equation}
is the probability mass function in $E$ that is closest to $p\left(x\right)$
in relative entropy. This is the conditional limit distribution with
respect to $p\left(x\right)$ and $E$. If, in addition, the set $E$
is the closure of its interior, then
\begin{equation}
\lim_{N\rightarrow\infty}\frac{1}{N}\log p^{N}\left(E\right)=-D\left(q^{\star}\left(X\right)\parallel p\left(X\right)\right).\label{eq:Sanov}
\end{equation}

The conditional limit theorem \cite[Theorem 11.6.2]{key-10} strengthens
the Sanov theorem by showing that with high probability the types
of the sequences which belong to $E$ are very close to $q^{\star}\left(x\right)$
in divergence, and the marginals tend towards $q^{\star}\left(x\right)$
when $N\rightarrow\infty$. The theorem is formulated as follows.
Let $E$ be a closed convex subset of $\mathcal{P}$ and let $p\left(x\right)$
be a probability mass function not in $E$. Let $X_{1},\ldots,X_{N}$
be discrete random variables drawn i.i.d. according to $p\left(x\right)$.
Let $q^{\star}\left(x\right)$ achieve $\min_{P\left(x\right)\in E}D\left(P\left(X\right)\parallel p\left(X\right)\right)$.
Then, 
\begin{equation}
\lim_{N\rightarrow\infty}\Pr\left(X_{k}=a\mid P_{\boldsymbol{X}}\in E\right)=q^{\star}\left(a\right),
\end{equation}
where $a\in\mathcal{X}$ and $1\leq k\leq N$.

\section{\label{sec:Subcode-construction}Shaped code construction}

As stated in the introduction, the goal is approaching the channel
capacity subject to a set of single letter constraints. Maximizing
the average mutual information subject to these constraints obtains
the maximal achievable transmission rate. 

Let $X$ and $Y$ be the input and output of a memoryless channel
with probability mass function $p\left(y\mid x\right)$. Let $E$
be a set of probability mass functions subjected to a set of single
letter constraints, given by
\begin{equation}
E=\left\{ P:\sum_{x\in\mathcal{X}}P\left(x\right)\varphi_{l}\left(x\right)\leq\beta_{l}\right\} ,\label{eq:E set}
\end{equation}
for $l=0,\ldots,L-1$, where $E$ is the closure of its interior,
$\varphi_{l}\left(x\right)$ are bounded constraint functions and
$\beta_{l}$ are constraints. 

It is well-known that i.i.d. input maximizes the average mutual information
subject to single letter constraints. Hence, the channel capacity
is
\begin{equation}
C_{E}=\max_{P\left(x\right)\in E}I_{P}\left(X;Y\right),
\end{equation}
and the probability mass function which achieves this capacity is
\begin{equation}
\hat{q}\left(x\right)=\arg\max_{P\left(x\right)\in E}I_{P}\left(X;Y\right).
\end{equation}

Let $R_{q}<C_{E}$ designate the desirable transmission rate. The
first method to create a random codebook is directly generate $2^{NR_{q}}$
codewords according to an i.i.d. probability mass function $q\left(x\right)$,
such that $R_{q}<I_{q}\left(X;Y\right)$. We refer to this method
as \emph{code construction I} with respect to $q\left(x\right)$.
The maximal reliable transmission rate can obviously be achieved using
code construction \emph{I} and $\hat{q}\left(x\right)$. 

Generating a practical good codebook with a specific probability mass
function is not trivial. Thus, the shaping method can be used. In
this method, a large codebook is generated according to an i.i.d.
uniform probability mass function and then codewords which satisfy
a set of constraints are chosen, creating the shaped subcode. 

We generalize this method, by allowing the large codebook to be generated
according to any i.i.d. probability mass function $p\left(x\right)$.
We refer to this method as \emph{code construction II} with respect
to $p\left(x\right)$ and $E$. We are motivated to analyze code construction
\emph{II}, since good practical codes for uniform probability mass
function are readily available, and codeword selection might be a
simple procedure for a properly structured code. 

More formally, let $\mathcal{C}$ be a random codebook with $2^{NR}$
codewords, referred to as the large codebook. Let $R_{s}$ be the
shaping rate. Let
\begin{equation}
R_{q}=R-R_{s}
\end{equation}
 be the shaped code rate. The large codebook is the union of $2^{NR_{q}}$
random sets, with $2^{NR_{s}}$ codewords in each random set. Each
codeword with length $N$ is generated according to i.i.d. probability
mass function $p\left(x\right)$ on $\mathcal{X}$. A message $\boldsymbol{m}$
to be transmitted chooses the random set. Out of this random set a
single codeword $\boldsymbol{X}=\left(X_{0},\ldots,X_{N-1}\right)\in\mathcal{C}$
is selected such that it satisfies the constraints
\begin{align}
 & \frac{1}{N}\sum_{i=0}^{N-1}\varphi_{l}\left(X_{i}\right)\leq\beta_{l}\Leftrightarrow\nonumber \\
 & \sum_{x\in\mathcal{X}}P_{\boldsymbol{X}}\left(x\right)\varphi_{l}\left(x\right)\leq\beta_{l}\Leftrightarrow\nonumber \\
 & P_{\boldsymbol{X}}\in E\cap\mathcal{P}_{N},\label{eq:Constraints}
\end{align}
where $P_{\boldsymbol{X}}$ is type of $\boldsymbol{X}$. A simple
illustration of the selection process can be seen in Fig. \ref{fig:Random-subcode-selection}.

The following are examples of constraints and selection metrics. The
first is the average power constraint with power limitation of $\beta_{0}$
\begin{equation}
\sum_{x\in\mathcal{X}}P\left(x\right)\left|x\right|^{2}\leq\beta_{0},
\end{equation}
using $\varphi_{0}\left(x\right)=\left|x\right|^{2}$. The selected
codewords have to satisfy $\frac{1}{N}\sum_{i=0}^{N-1}\left|X_{i}\right|^{2}\leq\beta_{0}$
to ensure that the codebook satisfies the constraints in average.
Another example is the hamming power constraint, where the number
of ones ('1') is limited to $\beta_{0}$, i.e., 
\begin{equation}
\sum_{x=0,1}P\left(x\right)x\leq\beta_{0},
\end{equation}
using $\varphi_{0}\left(x\right)=x$. In this case the selected codewords
have to satisfy $\frac{1}{N}\sum_{i=0}^{N-1}X_{i}\leq\beta_{0}$. 

For every constraint and constraint function $\varphi\left(\cdot\right)$,
the shaping rate $R_{s}$ is chosen such that a random set includes
at least one suitable codeword with probability approaching 1. The
union of the selected codewords is the shaped code $\mathcal{C}_{s}$.
Since each selected codeword $\boldsymbol{X}\in\mathcal{C}_{s}$ satisfies
the constraints, the shaped codebook satisfies the constraints in
average as required by (\ref{eq:E set}).

Recall that this code construction provides us with a framework for
approximating the performance of practical shaping schemes using information
theoretic tools. Thus, the questions in hand are how to choose the
shaping rate $R_{s}$ to find such non-typical codewords, and what
are the achievable rates of this codebook construction. In particular,
when using the matched decoder which is aware of the selection process
and the constraints, and the mismatched decoders which is aware only
of the large codebook.

The next section is dedicated to the summary of the main theorems,
which provides a full overview of the main results of this paper.
After which the reader can safely proceed to Section \ref{sec:Special-cases}
for special case and results.

\subsection{Summary of the Main Results }

Let the shaping code construction be code construction \emph{II}.
Let $R_{s}$ be the shaping rate. Given the set of constraints $E$,
let 
\begin{equation}
q^{\star}\left(x\right)=\arg\min_{P\left(x\right)\in E}D\left(P\left(X\right)\parallel p\left(X\right)\right)
\end{equation}
be the conditional limit distribution. Thus, for 
\begin{equation}
R_{s}>D\left(q^{\star}\left(X\right)\parallel p\left(X\right)\right)
\end{equation}
the probability that a random set of size $2^{NR_{s}}$ has at least
one codeword which satisfies the constraints approaches 1. Furthermore,
the probability that each random set includes at least one such codeword,
which is the probability that the shaped codebook exists, also approaches
1.

Let $R_{M}$ be the maximal achievable rate under the constraints
in $E$, code construction \emph{II} using $E$ and $p\left(x\right)$,
and the matched decoder. Then, 
\begin{equation}
R_{M}<\min\left\{ H_{p}\left(X\right)-D(q^{\star}\left(X\right)\parallel p\left(X\right)),I_{q^{\star}}\left(X;Y\right)\right\} 
\end{equation}
is the achievable rate. This results from the modified AEP theorem
(Theorem \ref{thm:Joint-AEP-CCII}), developed in this paper with
respect to code construction \emph{II}. 

For the mismatched decoder, let $R_{G}$ be the achievable rate resulting
from the modified development of the Gallager error exponent, which
suites code construction \emph{II}. Thus, for the rate
\begin{align}
 & R_{G}<\min\left\{ H_{p}\left(X\right)-D(q^{\star}\left(X\right)\parallel p\left(X\right)),\right.\nonumber \\
 & \left.I_{q^{\star}}\left(X;Y\right)+D\left(q^{\star}\left(Y\right)\parallel p\left(Y\right)\right)-D(q^{\star}\left(X\right)\parallel p\left(X\right))\right\} 
\end{align}
 the average error probability approaches 0, i.e., this is an achievable
transmission rate. Furthermore, let $R_{MJT}$ be the achievable rate
resulting from the modified joint typicality analysis given in Section
\ref{sec:Joint-Typicality-Decoder}. This rate is achievable in special
cases only, in particular when the conditions of Lemma \ref{lem:First error event}
are satisfied. Thus in those special cases,
\begin{align}
 & R_{MJT}<\min\left\{ H_{p}\left(X\right)-D(q^{\star}\left(X\right)\parallel p\left(X\right)),\right.\nonumber \\
 & \left.D\left(P^{\star}\left(X,Y\right)||p\left(X\right)q^{\star}\left(Y\right)\right)-D(q^{\star}\left(X\right)\parallel p\left(X\right))\right\} 
\end{align}
is an achievable rate, where the distribution $P^{\star}\left(x,y\right)$
is given in (\ref{eq:Pstar-MJT}). Under the conditions of Lemma \ref{lem:First error event},
the maximum between the two bounds is the achievable rate. Otherwise,
only the Gallager bound can be used.

The remaining of the paper is dedicated to elaborated presentation
of these results, formulating theorems and proofs, and analyzing interesting
and practical special cases.

\subsection{Shaping rate}

Thus, this section is dedicated to choosing $R_{s}$. This choice
has to ensure that the large codebook has the right amount of codewords
to support the transmission rate $R_{q}$, i.e., that each random
set of size $2^{NR_{s}}$ has at least one codeword which satisfies
the constraints with very high probability. In Theorem \ref{thm:Number of NT codewords},
we show that in the limit, this probability approaches 1 if $R_{s}>D\left(q^{\star}\left(X\right)\parallel p\left(X\right)\right),$
where $q^{\star}\left(x\right)=\arg\min_{P\left(x\right)\in E}D\left(P\left(X\right)\parallel p\left(X\right)\right)$.
The proof highly relies on the method of types and the Sanov theorem.
We also prove that (\ref{eq:RsPlusdelta}) guarantees a minimum on
$R_{s}$, such that the shaped codebook (using code construction \emph{II})
exists with very high probability. This is one of the main results
of this paper.
\begin{thm}
\label{thm:Number of NT codewords} Let $R_{s}$ be the shaping rate
in code construction II. Let \emph{$E$} be a set of probability mass
functions, which is a closure of its interior. Let $P_{s}$ designate
the probability of finding at least one codeword which satisfies the
constraints (type belongs to $E$) in a random set of size $2^{NR_{s}}$,
generated according to i.i.d. probability mass function $p\left(x\right)$.
Let \textup{$q^{\star}\left(x\right)$} be the conditional limit probability
mass function with respect to $p\left(x\right)$ and $E$, which achieves
$\min_{P\left(x\right)\in E}D\left(P\left(X\right)\parallel p\left(X\right)\right)$.

Then:\end{thm}
\begin{enumerate}
\item \emph{For any $\gamma$ there exists $N_{\gamma}$, such that for
$N>N_{\gamma}$
\begin{align}
 & P_{s}\geq\nonumber \\
 & 1-\left(1-\frac{2^{-N\left(D\left(q^{\star}\left(X\right)\parallel p\left(X\right)\right)+\gamma\right)}}{\left(N+1\right)^{\left|\mathcal{X}\right|}}\right)^{2^{NR_{s}}}.
\end{align}
}
\item \emph{For any $\delta$ and $N\rightarrow\infty$, such that 
\begin{eqnarray}
R_{s} & = & D\left(q^{\star}\left(X\right)\parallel p\left(X\right)\right)+\delta,\label{eq:RsPlusdelta}
\end{eqnarray}
 $P_{s}\rightarrow1$}.
\item \emph{For any $\delta$ and any finite $\rho$, such that (\ref{eq:RsPlusdelta})
is satisfied,
\begin{equation}
P_{\rho}\triangleq\left(P_{s}\right)^{2^{\rho\cdot N}}\rightarrow1
\end{equation}
for $N\rightarrow\infty$.}\end{enumerate}
\begin{IEEEproof}
The probability that a codeword satisfies the constraints $p^{N}\left(E\right)$,
is given by the Sanov theorem (\ref{eq:p(N)E}). Let $\mathcal{P}_{N}$
be the set of types with denominator $N$. Since $E$ is the closure
of its interior, it follows that $E\cap\mathcal{P}_{N}$ is nonempty
for $N>N_{0}$. Hence, we can find a sequence of probability mass
functions $Q_{N}$ such that $Q_{N}\in E\cap\mathcal{P}_{N}$ and
\begin{equation}
D\left(Q_{N}\parallel p\left(X\right)\right)\rightarrow D\left(q^{\star}\left(X\right)\parallel p\left(X\right)\right),
\end{equation}
since by definition $q^{\star}\left(x\right)\in E$. This means that
for any $\gamma$ there exists $N_{\gamma}>N_{0}$, such that
\begin{equation}
D\left(Q_{N}\parallel p\left(X\right)\right)<D\left(q^{\star}\left(X\right)\parallel p\left(X\right)\right)+\gamma.\label{eq:seq divergence}
\end{equation}
Thus, for $N>N_{\gamma}$ 
\begin{align}
 & p^{N}\left(E\right)\geq p^{N}\left(T\left(Q_{N}\right)\right)\geq\nonumber \\
 & \frac{1}{\left(N+1\right)^{\left|\mathcal{X}\right|}}2^{-ND\left(Q_{N}\parallel p\left(x\right)\right)}\geq\nonumber \\
 & \frac{1}{\left(N+1\right)^{\left|\mathcal{X}\right|}}2^{-N\left(D\left(q^{\star}\left(X\right)\parallel p\left(X\right)\right)+\gamma\right)},\label{eq:bound on PN(E)}
\end{align}
where the inequalities follow from (\ref{eq:Type probability}) and
(\ref{eq:seq divergence}). It yields that for $N>N_{\gamma}$, the
probability that none of the codewords within a random set satisfies
the constraints is
\begin{align}
 & P_{f}=\nonumber \\
 & \left(1-p^{N}\left(E\right)\right)^{2^{NR_{s}}}\leq\nonumber \\
 & \left(1-\frac{2^{-N\left(D\left(q^{\star}\left(X\right)\parallel p\left(X\right)\right)+\gamma\right)}}{\left(N+1\right)^{\left|\mathcal{X}\right|}}\right)^{2^{NR_{s}}}.
\end{align}
Since $P_{s}=1-P_{f}$, this concludes the first part.

For any $\delta$ and the choice $R_{s}=D\left(q^{\star}\left(X\right)\parallel p\left(X\right)\right)+\delta$,
\begin{align}
 & \log P_{f}=\nonumber \\
 & 2^{NR_{s}}\log\left(1-p^{N}\left(E\right)\right)\leq\nonumber \\
 & -2^{N\left(D\left(q^{\star}\left(X\right)\parallel p\left(X\right)\right)+\delta\right)}p^{N}\left(E\right)=\nonumber \\
 & -2^{N\left(D\left(q^{\star}\left(X\right)\parallel p\left(X\right)\right)+\delta+\frac{1}{N}\log p^{N}\left(E\right)\right)}.
\end{align}
Furthermore, according to (\ref{eq:Sanov}) for every $\delta$ there
exists $\zeta<\delta$ and $N_{\zeta}$, such that for $N>N_{\zeta}$
\begin{equation}
\frac{1}{N}\log p^{N}\left(E\right)\geq-D\left(q^{\star}\left(X\right)\parallel p\left(X\right)\right)-\zeta.
\end{equation}
Hence for $N>N_{\zeta}$,
\begin{equation}
\log P_{f}\leq-2^{N\left(\delta-\zeta\right)}
\end{equation}
and for $N\rightarrow\infty$, $\log P_{f}\rightarrow-\infty$. Thus
$P_{f}\rightarrow0$ and $P_{s}\rightarrow1$. This concludes the
second part.

Furthermore,

\begin{eqnarray}
P_{\rho} & = & \left(P_{s}\right)^{2^{\rho\cdot N}}\nonumber \\
 & = & \left(1-P_{f}\right)^{2^{\rho\cdot N}}\nonumber \\
 & = & \left[\left(1-P_{f}\right)^{1/P_{f}}\right]^{2^{\rho\cdot N}P_{f}}.
\end{eqnarray}
For $N\rightarrow\infty$, since $P_{f}\rightarrow0$, it yields that
$\left(1-P_{f}\right)^{1/P_{f}}\rightarrow1/e$. For $N>N_{\zeta}$,
\begin{eqnarray}
\log2^{NR_{q}}P_{f} & = & \rho\cdot N+\log P_{f}\nonumber \\
 & \leq & \rho\cdot N-2^{N\left(\delta-\zeta\right)}.
\end{eqnarray}
This yields that for $N\rightarrow\infty$, $\log2^{\rho\cdot N}P_{f}\rightarrow-\infty$
and therefore $2^{\rho\cdot N}P_{f}\rightarrow0$. Hence, $P_{\rho}\rightarrow1$. 
\end{IEEEproof}
Thus, we found the shaping rate which ensures that a random set has
at least one suitable codeword. Since the random sets are drawn independently,
the probability that each one of the random sets has at least one
suitable codeword is $P_{\mathcal{C}_{s}}=\left(P_{s}\right)^{2^{NR_{q}}}$.
Using the last theorem with $\rho=R_{q}$, we conclude that $P_{\mathcal{C}_{s}}\rightarrow1$,
i.e., the shaped codebook (code construction \emph{II}) exists with
high probability.

\subsection{\label{sub:Number-of-codewords}Number of codewords}

In the previous section we have shown that using a large code with
code rate $R$ and redundancy shaping rate $R_{s}$ a shaped subcode
can be obtained. This subcode consists of $2^{N\left(R-R_{s}\right)}$
codewords which satisfy the constraints with probability tending towards
1. Hence, the maximal number of codewords in the subcode is
\begin{eqnarray}
R_{q} & < & H_{p}\left(X\right)-D(q^{\star}\left(X\right)\parallel p\left(X\right))\label{eq:maximal number of code words}\\
 & \leq & H_{q^{\star}}\left(X\right).\nonumber 
\end{eqnarray}
The last inequality follows since the number of chosen codewords can
not exceed the size of the typical set of $q^{\star}\left(x\right)$.
This property holds as long as $p\left(x\right)\notin E$. Whereas
for $p\left(x\right)\in E$, the codewords of the large code satisfy
the constraints and $R_{s}=0$. 

We conclude that the achievable rate would be the minimum between
the number of codewords and the rate which achieves average error
probability tending towards zero for $N\rightarrow\infty$ . To establish
this rate we analyze the performance of several decoders, which are
presented next.

\subsection{Decoder types}

Following our codebook construction method, we consider two types
of decoders. The first is the matched decoder, which is aware of the
codebook set $\mathcal{C}_{s}$ and performs optimal decoding. The
second decoder is the mismatched decoder, which is only aware of the
large code $\mathcal{C}$ and performs optimal decoding as if all
codewords of $\mathcal{C}$ might have been transmitted. This decoder
is suboptimal, but occasionally more practical than the matched decoder.
In the practical special case where the large codebook is uniform
code, the mismatched decoder is readily available while the adaptation
to the set $\mathcal{C}_{s}$ might be extremely complex. 

The achievable rate of code construction \emph{I} is known. It is
the mutual information related to the probability mass function which
generated the codebook and the probability mass function of the channel.
The achievable rate of code construction \emph{II} decoded with the
matched decoder, however, is an open question. It is the objective
of Section \ref{sec:Matched-decoder} to show that the achievable
rate, in this case, is bounded by $I_{q^{\star}}\left(X;Y\right)$.

On first look, mismatch performance analysis seems as a trivial exercise.
The transmitted codeword is part of the large codebook and therefore
the large code performance can be used. This is the naive approach.
This type of assumption was taken (without being mentioned) by Forney
\cite{key-1}, using a decoder which is unaware of the selection process
and stating that the performance and complexity of their decoder are
not affected by shaping. A more careful inspection, however, reveals
that the large code performance assumes the transmission of the typical
codewords, while we transmit only non-typical codewords. These codewords
may have worse distance spectrum towards their neighbors. 

The average error probability of the large codebook tends towards
zero when $R<I_{p}\left(X;Y\right)$. Thus, the bound on $R_{q}$
using the naive approach is 
\begin{eqnarray}
R_{n}\triangleq R_{q} & < & I_{p}\left(X;Y\right)-R_{s}\nonumber \\
 & = & I_{p}\left(X;Y\right)-D\left(q^{\star}\left(X\right)\parallel p\left(X\right)\right).\label{eq:Naive approach rate.}
\end{eqnarray}
This bound is achievable in special cases only, for example the BSC
channel, as will be shown. This is also correct for codes (bad codes),
which operate at working point that justifies the approximation of
minimum distance, and this justifies the analysis of Forney. Another
interesting special case is a large code construction in which all
words have the same distance spectrum. Obviously in such case the
selection of specific codewords will not make a difference. For the
AWGN, geometrically uniform (GU) codes (e.g. linear codes over rings)
are known only to MPSK constellation. This raises an interesting open
question: is the mismatched approximation accurate for MPSK? We know
that GU codes generate good codes, and we can use them for the large
uniform codebook. If the conditional limit theorem holds for GU codes
over MPSK, then we can also know the choice of $R_{s}$ that will
generate enough codewords that satisfy the constraints.

\subsection{Related practical schemes}

The described random shaping code construction method highly resembles
existing practical schemes. Since well-known good codes (e.g. convolutional
codes, LDPC, block codes) typically generate uniform probability mass
function, combined with a selection process the outcome is a shaped
subcode. Our results are bounds on the performance of random constructions
which approximate practical systems, which are approached using uniform
i.i.d. probability mass function, the right shaping rate $R_{s}$
and sufficiently large $N$. Next we present two examples of such
practical schemes.

\subsubsection{TS - Sign Bit Shaping}

The trellis shaping \cite{key-1}, is a sequence oriented approach.
\begin{figure}
\centering{}\includegraphics[width=8cm,height=8cm,keepaspectratio]{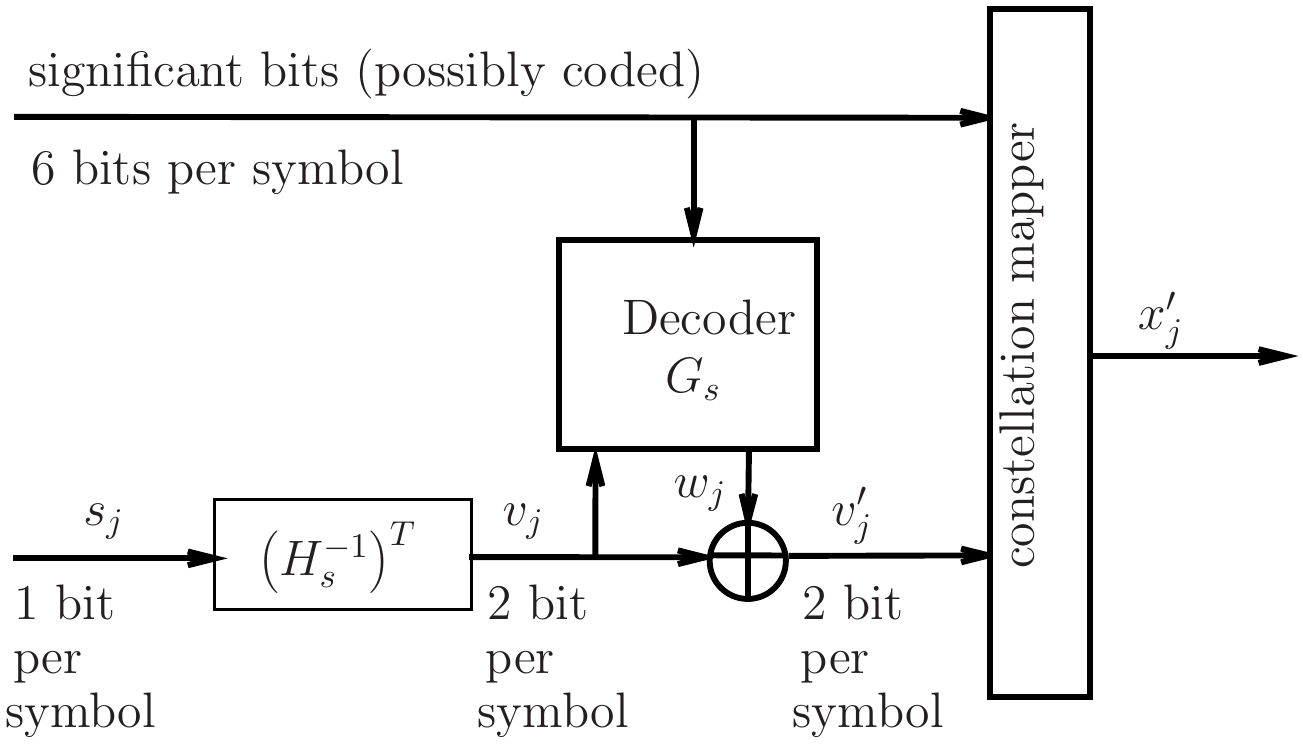}\caption{\label{fig:Bit-sign-shaping}Bit Sign Shaping}
\end{figure}
The main idea is to use a trellis shaping code to create equivalence
classes of coded sequences and then perform the shaping operation
by choosing the minimum power sequence to be transmitted within each
equivalence class. Sign Bit Shaping (SBS), is a special case of TS.
In SBS, two bits per symbol $v_{i}=(v_{i_{1}},v_{i_{2}})$ are modified
by rate 1/2 trellis shaping code, having generating matrix $\boldsymbol{G}_{s}$
and parity check matrix $\boldsymbol{H}_{s}$. 

Let $N$ be the codeword length (symbols), $M$ be the number of bits
represented by each symbol and $f\left(\cdot\right)$ be a mapping
function to the euclidean space. Each symbol has $M-2$ unshaped bits,
possibly coded by a channel code $C_{c}$ with rate $R_{c}$, and
two shaped sign bits, as shown in Fig. \ref{fig:Bit-sign-shaping}.
The sign bit sequence is 
\begin{equation}
\boldsymbol{v}=\boldsymbol{s}\cdot\left(\boldsymbol{H}_{s}^{-1}\right)^{T},
\end{equation}
where $\boldsymbol{s}$ are $N$ information bits. The trellis decoder
creates an equivalence class
\begin{equation}
\left\{ \boldsymbol{v}^{\prime}\right\} =\boldsymbol{v}+\left\{ \boldsymbol{w}\right\} ,
\end{equation}
where $\left\{ \boldsymbol{w}\right\} =\boldsymbol{d}\boldsymbol{G}_{s}$
and $\boldsymbol{d}\in\left\{ 0,1\right\} ^{N}$ represents all binary
sequences of length $N$.

The bits are mapped into equivalence class of symbols $\left\{ \boldsymbol{x}^{\prime}\right\} =f\left(\left\{ \boldsymbol{v}^{\prime}\right\} ,\boldsymbol{u}\right)$.
In this scheme the goal is power minimization; therefore the trellis
decoder transmits the codeword with the minimal energy within the
equivalence class, i.e.,
\begin{equation}
\boldsymbol{x}=\arg\min_{\left\{ \boldsymbol{x}^{\prime}\right\} }\left\{ \left|\boldsymbol{x}^{\prime}\right|^{2}\right\} ,
\end{equation}
or the minimal average energy
\begin{equation}
\boldsymbol{x}=\arg\min_{\left\{ \boldsymbol{x}^{\prime}\right\} }\left\{ \frac{1}{N}\sum_{i=0}^{N-1}\left|x_{i}^{\prime}\right|^{2}\right\} .\label{eq:avg eng}
\end{equation}
Therefore, the total information rate of this scheme is 
\begin{equation}
R_{q}=R_{c}\left(M-2\right)+1
\end{equation}
bits per symbol and the shaping rate is $R_{s}=1$. 

The decoding process is within the large codebook, which includes
all the equivalence classes. Once the codeword $\hat{\boldsymbol{x}}$
has been decoded, the demapper extracts the shaped bits and the only
additional operation required is reconstruction the shaped bits by
\begin{equation}
\hat{\boldsymbol{s}}=\boldsymbol{H}_{s}\left(\boldsymbol{v}+\boldsymbol{d}\boldsymbol{G}_{s}\right)=\boldsymbol{s}.
\end{equation}
This operation is error free in case the entire codeword was decoded
correctly ($\boldsymbol{x}=\hat{\boldsymbol{x}}$) and since $\boldsymbol{H}_{s}\cdot\boldsymbol{G}_{s}=0$.

The selection process in this scheme is to choose the codeword with
the minimal average power in each equivalence class, as given in (\ref{eq:avg eng}).
For very large $N$, we claim that this selection is equivalent to
selecting a sequence which satisfies the power constraint
\begin{equation}
\boldsymbol{x}:\;\frac{1}{N}\sum_{i=0}^{N-1}\left|x_{i}^{\prime}\right|^{2}\leq\beta_{0}.
\end{equation}
Since the shaping rate $R_{s}$ is set by the shaping code $\boldsymbol{G}_{s}$
and the constraint metric $\varphi\left(x\right)=\left|x\right|^{2}$
is given, there exists minimal $\beta_{0}$ such that each equivalence
class has at least one such codeword. This is dual to choosing a constraint
and finding the minimal $R_{s}$ for this constraint. The constraint
in turn, defines the probability mass function of the selected codewords.
Recall that these sequences are very close to $q^{\star}\left(x\right)$
in distribution. The minimum average energy codeword also satisfies
the constraint, thus it is one of the codewords in the previous method.
Although, the chosen codewords might differ, using the different methods,
the outcome is the same. A code which satisfies the constraint and
is very close to $q^{\star}\left(x\right)$. Thus, choosing the minimum
and choosing the codeword satisfying the minimal $\beta_{0}$ is asymptotically
the same. Assuming that the large codebook is uniformly distributed,
$q^{\star}\left(x\right)$ is the entropy maximizing Maxwell-Boltzmann
probability mass function \cite{key-10}. 

Note that in case of several constraints the methods are not longer
equivalent since several constraints can not be represented by single
metric, which can be minimized using the TS decoder at the transmitter.

\subsubsection{Compound LDGM-LDPC}

Another example to the selection process is the scheme in \cite{key-7},
which was originally designed for dirty paper coding. This scheme
creates a shaping code $\Lambda_{s}\left(\boldsymbol{m}\right)$ for
each transmitted message $\boldsymbol{m}$ of length $NR$ using 
\begin{eqnarray}
\Lambda_{s}\left(\boldsymbol{m}\right) & = & \left\{ \boldsymbol{c}=\boldsymbol{b}\boldsymbol{G}|\boldsymbol{b}\in\left\{ 0,1\right\} ^{NR_{b}},\right.\nonumber \\
 &  & \left.\boldsymbol{b}\boldsymbol{H}_{p}^{T}=\boldsymbol{p},\,\boldsymbol{b}\boldsymbol{H}_{m}^{T}=\boldsymbol{m}\right\} ,
\end{eqnarray}
where $\boldsymbol{G}$ is a sparse generation matrix, $\boldsymbol{H}_{p}$
and $\boldsymbol{H}_{m}$ are sparse parity check matrices, and $\boldsymbol{p}$
is a fixed binary sequence of length $NR_{p}$. In this nested LDGM-LDPC
scheme the matrix $\boldsymbol{G}$ is designed to make the shaping
code $C_{s}\left(\boldsymbol{m}\right)$ a good source code and thus
suitable for average power shaping. The matrices $\boldsymbol{H}_{p}$
and $\boldsymbol{H}_{m}$ embed information and obtain good channel
code properties. The encoder, given the information massage $\boldsymbol{m}$,
performs a mean square error (MSE) quantization of a scaled interference
(according to the dirty paper scheme) using a lattice $\Lambda_{s}\left(\boldsymbol{m}\right)+2\mathbb{Z}^{N}$
and transmits the quantization error. A special use of this scheme
is the quantization of a zero interference, which results in transmitting
a minimum energy codeword in the shaping code for each message. The
union of the shaping codes is the large codebook and the subcode $C_{s}$
is the union of all minimum energy codewords selected from each shaping
code $\Lambda_{s}\left(\boldsymbol{m}\right)$. The information rate
using this scheme is $R$ and the shaping rate can be obtained from
the size of the shaping code $\left|\Lambda_{s}\left(\boldsymbol{m}\right)\right|=2^{NR_{s}}$,
i.e., 
\begin{equation}
R_{s}=R_{b}-R_{p}-R_{m}.
\end{equation}
The selection process is the same as in the SBS scheme, therefore
the same conclusions apply to this scheme.

\section{\label{sec:Matched-decoder}Matched decoder: Joint Typicality Decoding}

In previous sections, we discussed code construction \emph{II} as
a practical method to obtain a codebook which satisfies constraints.
The objective of this section is showing the achievable rate of this
code construction method and the matched decoder, with respect to
the large codebook probability mass function $p\left(x\right)$ and
the set of constraints $E$. Furthermore, we bound the $N$-dimensional
distribution of the codewords, using the conditional limit distribution.
These are new results, since the conditional limit theorem treated
the distribution of the marginals of the selected codewords, but not
the entropy rate nor the average mutual information when transmitted
over a channel.

\subsection{Codeword Distribution}

Let $p\left(x\right)$ be i.i.d. probability mass function on $\mathcal{X}$
and let $E$ be a closed convex set such that $p\left(x\right)\notin E$.
Let $q^{\star}\left(x\right)$ be the conditional limit probability
mass function, which achieves $\min_{P\left(x\right)\in E}D\left(P\left(X\right)\parallel p\left(X\right)\right)$.
Let us have a standard random codebook (code construction \emph{I})
which uses this distribution to generate the codewords. Thus, 
\begin{equation}
q^{\star}\left(\boldsymbol{x}\right)=\prod_{i=1}^{N}q^{\star}\left(x_{i}\right)
\end{equation}
is the probability to generate the codeword $\boldsymbol{x}$ and
\begin{eqnarray}
q^{\star}\left(\boldsymbol{y}\right) & = & \sum_{\boldsymbol{x}\in\mathcal{X}^{N}}\prod_{i=1}^{N}q^{\star}\left(x_{i}\right)p\left(y_{i}\mid x_{i}\right)
\end{eqnarray}
is the probability to generate the channel output $\boldsymbol{y}$.
Let $A_{\epsilon,q^{\star}\left(X,Y\right)}^{(N)}$ designate the
jointly typical set in definition \ref{A_pxy}, and let the decoder
be the joint typicality decoder with respect to\emph{ }the i.i.d.
probability mass function\emph{ }$q^{\star}\left(x,y\right)=q^{\star}\left(x\right)p\left(y\mid x\right)$.
Let $\Pr\left\{ \boldsymbol{x}\mid CC_{II}\right\} $, $\Pr\left\{ \boldsymbol{y}\mid CC_{II}\right\} $
designate the probability to generate the codeword $\boldsymbol{x}$
and the channel output $\boldsymbol{y}$, respectively, using code
construction \emph{II}. Let $R_{s}>D\left(q^{\star}\left(X\right)\parallel p\left(X\right)\right)$.
The next two lemmas upper bound $\Pr\left\{ \boldsymbol{x}\mid CC_{II}\right\} $,
$\Pr\left\{ \boldsymbol{y}\mid CC_{II}\right\} $ using $q^{\star}\left(\boldsymbol{x}\right)$
and $q^{\star}\left(\boldsymbol{y}\right)$. These bounds are later
used to derive the modified AEP theorem for code construction \emph{II}
(Theorem \ref{thm:Joint-AEP-CCII}). The AEP, which defines the average
probability of the decoding error, allows to deduct the achievable
rate. 
\begin{lem}
\label{lem:bounding CCII with CCI}For every $\delta$ there exists
$N_{\delta}$ such that for $N>N_{\delta}$
\begin{equation}
\Pr\left\{ \boldsymbol{x}\mid CC_{II}\right\} \leq\left(N+1\right)^{\left|\mathcal{X}\right|}2^{N\delta}q^{\star}\left(\boldsymbol{x}\right).
\end{equation}
\end{lem}
\begin{IEEEproof}
Since $p\left(x\right)$ is i.i.d. probability mass function on $\mathcal{X}$,
the probability to generate a codeword $\boldsymbol{x}$ from probability
mass function $p\left(x\right)$ is 
\begin{equation}
p^{N}\left(\boldsymbol{x}\right)=2^{-N\left(H\left(Q_{\boldsymbol{x}}\right)+D\left(Q_{\boldsymbol{x}}\parallel p\left(X\right)\right)\right)},
\end{equation}
where $Q_{\boldsymbol{x}}$ is the type of $\boldsymbol{x}$. The
probability to generate the codeword $\boldsymbol{x}$, using code
construction \emph{II}, is 
\begin{equation}
\Pr\left\{ \boldsymbol{x}\mid CC_{II}\right\} =\left\{ \begin{array}{cc}
\frac{2^{-N\left(H\left(Q_{\boldsymbol{x}}\right)+D\left(Q_{\boldsymbol{x}}\parallel p\left(X\right)\right)\right)}}{p^{N}\left(E\right)}, & Q_{\boldsymbol{x}}\in E,\\
0, & Q_{\boldsymbol{x}}\notin E,
\end{array}\right.
\end{equation}
where $p^{N}\left(E\right)$ is the probability of the event $Q_{\boldsymbol{X}}\in E$,
when generating from i.i.d. $p\left(x\right)$. Since $E$ is a closed
convex set, (\ref{eq:bound on PN(E)}) applies. Furthermore, by Pythagorean
theorem for divergence \cite[Theorem 11.6.1]{key-10}, every $P\in E$
satisfies
\begin{align}
 & D\left(P\parallel p\left(X\right)\right)\geq D\left(P\parallel q^{\star}\left(X\right)\right)+D\left(q^{\star}\left(X\right)\parallel p\left(X\right)\right),\label{eq:Pythagorean}
\end{align}
if $E$ is a closed convex set and $p\left(x\right)\notin E$. Thus,
\begin{alignat}{1}
 & \Pr\left\{ \boldsymbol{x}\mid CC_{II}\right\} \leq\nonumber \\
 & \frac{2^{-N\left(H\left(Q_{\boldsymbol{x}}\right)+D\left(Q_{\boldsymbol{x}}\parallel q^{\star}\left(X\right)\right)+D\left(q^{\star}\left(X\right)\parallel p\left(X\right)\right)\right)}}{p^{N}\left(E\right)}\leq\nonumber \\
 & \left(N+1\right)^{\left|\mathcal{X}\right|}2^{-N\left(H\left(Q_{\boldsymbol{x}}\right)+D\left(Q_{\boldsymbol{x}}\parallel q^{\star}\left(X\right)\right)-\delta\right)},
\end{alignat}
using (\ref{eq:bound on PN(E)}) and (\ref{eq:Pythagorean}).

The probability to generate the codeword $\boldsymbol{x}$ with respect
to i.i.d. probability mass function $q^{\star}\left(x\right)$ using
code construction \emph{I} (according to (\ref{eq:seq. probability}))
is 
\begin{equation}
q^{\star}\left(\boldsymbol{x}\right)=2^{-N\left(H\left(Q_{\boldsymbol{x}}\right)+D\left(Q_{\boldsymbol{x}}\parallel q^{\star}\left(X\right)\right)\right)}.
\end{equation}
Hence, we conclude that $\forall\delta$ there exists $N_{\delta}$
such that for $N>N_{\delta}$ 
\begin{equation}
\Pr\left\{ \boldsymbol{x}\mid CC_{II}\right\} \leq\left(N+1\right)^{\left|X\right|}2^{N\delta}q^{\star}\left(\boldsymbol{x}\right).
\end{equation}
\end{IEEEproof}
\begin{lem}
\label{lem:bounding CCII with CCI Y}For every $\delta$ there exists
$N_{\delta}$ such that for $N>N_{\delta}$
\begin{equation}
\Pr\left\{ \boldsymbol{y}\mid CC_{II}\right\} \leq\left(N+1\right)^{\left|\mathcal{X}\right|}2^{N\delta}q^{\star}\left(\boldsymbol{y}\right).
\end{equation}
\end{lem}
\begin{IEEEproof}
The probability of $\boldsymbol{y}$ is the sum of the joint probabilities
$\Pr\left\{ \boldsymbol{x},\boldsymbol{y}\right\} $ over all the
codewords $\boldsymbol{x}$. Thus, $\forall\delta$ there exists $N_{\delta}$
such that for $N>N_{\delta}$
\begin{alignat}{1}
 & \Pr\left\{ \boldsymbol{y}\mid CC_{II}\right\} =\nonumber \\
 & \sum_{\boldsymbol{x}:Q_{\boldsymbol{x}}\in E}\Pr\left\{ \boldsymbol{x}\mid CC_{II}\right\} p\left(\boldsymbol{y}\mid\boldsymbol{x}\right)\leq\nonumber \\
 & \sum_{\boldsymbol{x}:Q_{\boldsymbol{x}}\in E}\left(N+1\right)^{\left|\mathcal{X}\right|}2^{N\delta}q^{\star}\left(\boldsymbol{x}\right)p\left(\boldsymbol{y}\mid\boldsymbol{x}\right)\leq\nonumber \\
 & \sum_{\boldsymbol{x}\in\mathcal{X}^{N}}\left(N+1\right)^{\left|\mathcal{X}\right|}2^{N\delta}q^{\star}\left(\boldsymbol{x}\right)p\left(\boldsymbol{y}\mid\boldsymbol{x}\right)=\nonumber \\
 & \left(N+1\right)^{\left|\mathcal{X}\right|}2^{N\delta}q^{\star}\left(\boldsymbol{y}\right),
\end{alignat}
where the first inequality follows from Lemma \ref{lem:bounding CCII with CCI}.
\end{IEEEproof}

\subsection{Modified AEP}

One of the methods to obtain the achievable rate of a random codebook
ensemble which was drawn according to an i.i.d. probability mass function
is the channel coding theorem \cite[Theorem 7.7.1]{key-10}. It relies
on the joint AEP theorem \cite[Theorem 7.6.1]{key-10}. The AEP establishes
the probability that a codeword and a channel output are jointly typical
with respect to their joint probability mass function in two cases.
In case the output is a noisy version of the input and in case of
two independent variables. These two probabilities are essential to
analyze the average probability of error and deduct the achievable
rate such that the probability of error is tending towards zero when
the codewords length tends towards infinity. The decoder is the joint
typicality decoder which operates as follows. An error occurs if the
transmitted codeword is not jointly typical with the received sequence
or when more than one codeword is jointly typical. 

The AEP theorem is not suitable for code construction \emph{II}, since
the codewords are generated by a selection process and not randomly
drawn according to an i.i.d. distribution. Nevertheless, we know that
their probability mass function converges in a sense to the i.i.d.
conditional limit probability mass function. Thus, we can use this
conditional limit probability mass function to redefine the decoder
and derive a modified AEP theorem (Theorem \ref{thm:Joint-AEP-CCII}).
This theorem is used to show one of the main results of this paper,
that (\ref{eq:Matched achievable rate}) is the achievable rate under
matched decoder. We conclude that the same rate can be achieved either
by code construction \emph{II} or by random i.i.d. generation according
to $q^{\star}\left(x\right)$, i.e. $I_{q^{\star}}\left(X;Y\right)$
if the large codebook is uniformly distributed. It follows that code
construction \emph{II} does not lose rate under matched decoding in
this case.
\begin{thm}
\label{thm:Joint-AEP-CCII}(Joint AEP for code construction II) Let
\textup{$\boldsymbol{X}$, }\textup{\emph{$\boldsymbol{Y}$}}\textup{
}\textup{\emph{be a codeword of length $N$ drawn by code construction
II with respect to $E$ and the output of memoryless channel with
probability mass function $p\left(y\mid x\right)$, respectively.
}}Let \textup{\emph{$\tilde{\boldsymbol{X}}$, $\tilde{\boldsymbol{Y}}$
be a codeword and a channel output generated by code construction
II with respect to $E$, which are independent.}} Let $A_{\epsilon,q^{\star}\left(X,Y\right)}^{(N)}$
designate the jointly typical set in definition \ref{A_pxy}, \textup{\emph{such
that}}\emph{ }$q^{\star}\left(x,y\right)=q^{\star}\left(x\right)p\left(y\mid x\right)$.\textup{\emph{
Then:}}\end{thm}
\begin{enumerate}
\item \emph{$\Pr\left\{ \left(\boldsymbol{X},\boldsymbol{Y}\right)\in A_{\epsilon,q^{\star}\left(X,Y\right)}^{(N)}\right\} \rightarrow1$
as $N\rightarrow\infty$.}
\item \emph{For every $\epsilon,\delta$, the exist $N_{\epsilon},N_{\delta}$
such that for $N\geq\max\left(N_{\delta},N_{\epsilon}\right)$
\begin{alignat}{1}
 & \Pr\left\{ \left(\tilde{\boldsymbol{X}},\tilde{\boldsymbol{Y}}\right)\in A_{\epsilon,q^{\star}\left(X,Y\right)}^{(N)}\right\} \leq\nonumber \\
 & \left(N+1\right)^{2\left|\mathcal{X}\right|}2^{-N\left(I_{q^{\star}}\left(X;Y\right)-3\epsilon-2\delta\right)}.
\end{alignat}
}\end{enumerate}
\begin{IEEEproof}
1) Since $q^{\star}(\boldsymbol{X})=\prod_{i=0}^{N-1}q^{\star}\left(X_{i}\right)$,
it yields that
\begin{align}
 & -\frac{1}{N}\log q^{\star}(\boldsymbol{X})=\nonumber \\
 & -\frac{1}{N}\sum_{i=0}^{N-1}\log q^{\star}\left(X_{i}\right)=\nonumber \\
 & -\sum_{x\in\mathcal{X}}Q_{\boldsymbol{X}}\left(x\right)\log q^{\star}(x)=\nonumber \\
 & H_{Q_{\boldsymbol{X}}}\left(X\right)+D\left(Q_{\boldsymbol{X}}\left(X\right)\parallel q^{\star}(X)\right),
\end{align}
where $Q_{\boldsymbol{X}}$ is the type of $\boldsymbol{X}$. Thus,
\begin{alignat}{1}
 & \left|-\frac{1}{N}\log q^{\star}(\boldsymbol{X})-H_{q^{\star}}\left(X\right)\right|=\nonumber \\
 & \left|H_{Q_{\boldsymbol{X}}}\left(X\right)-H_{q^{\star}}\left(X\right)+D\left(Q_{\boldsymbol{X}}\left(X\right)\parallel q^{\star}(X)\right)\right|\leq\nonumber \\
 & -\left\Vert Q_{\boldsymbol{X}}\left(X\right)-q^{\star}\left(X\right)\right\Vert _{1}\log\frac{\left\Vert Q_{\boldsymbol{X}}\left(X\right)-q^{\star}\left(X\right)\right\Vert _{1}}{\left|\mathcal{X}\right|}+\nonumber \\
 & +D\left(Q_{\boldsymbol{X}}\left(X\right)\parallel q^{\star}(X)\right),\label{eq:first AEP cond}
\end{alignat}
where the last inequality follows from \cite[Theorem 17.3.3]{key-10}
and $\left\Vert P_{1}-P_{2}\right\Vert _{1}$ is the $\mathcal{L}_{1}$
distance between probability mass functions. The conditional limit
theorem \cite[Theorem 11.6.2]{key-10} implies that $\forall\gamma_{0}$
there exists $N_{0}$ such that for $N>N_{0}$ 
\begin{equation}
\Pr\left\{ D\left(Q_{\boldsymbol{X}}\left(X\right)\parallel q^{\star}(X)\right)\geq\gamma_{0}\mid Q_{\boldsymbol{X}}\in E\right\} \leq\eta.\label{eq:Div Qx q star}
\end{equation}
Furthermore, \cite[Lemma 11.6.1]{key-10} yields that 
\begin{equation}
\frac{1}{2\ln2}\left\Vert Q_{\boldsymbol{X}}\left(X\right)-q^{\star}\left(X\right)\right\Vert _{1}^{2}\leq D\left(Q_{\boldsymbol{X}}\left(X\right)\parallel q^{\star}(X)\right),
\end{equation}
thus if $D\left(Q_{\boldsymbol{X}}\left(X\right)\parallel q^{\star}(X)\right)\leq\gamma_{0}$,
the $\mathcal{L}_{1}$ distance 
\begin{equation}
\left\Vert Q_{\boldsymbol{X}}\left(X\right)-q^{\star}\left(X\right)\right\Vert _{1}\leq\gamma_{1}=\sqrt{2\ln2\gamma_{0}}.
\end{equation}
Since $\forall\epsilon$ there exists $\gamma_{0}$ such that 
\begin{equation}
-\gamma_{1}\log\frac{\gamma_{1}}{\left|\mathcal{X}\right|}+\gamma_{0}\leq\epsilon,
\end{equation}
from (\ref{eq:first AEP cond}) and (\ref{eq:Div Qx q star}) it yields
that there exists $N_{x}$ such that for $N>N_{x}$ 
\begin{equation}
\Pr\left\{ \left|-\frac{1}{N}\log q^{\star}(\boldsymbol{X})-H_{q^{\star}}\left(X\right)\right|\geq\epsilon\right\} \leq\epsilon/3.
\end{equation}

In a similar manner, 
\begin{alignat}{1}
 & \left|-\frac{1}{N}\log q^{\star}(\boldsymbol{Y})-H_{q^{\star}}\left(Y\right)\right|\leq\nonumber \\
 & -\left\Vert Q_{\boldsymbol{Y}}\left(Y\right)-q^{\star}\left(Y\right)\right\Vert _{1}\log\frac{\left\Vert Q_{\boldsymbol{Y}}\left(Y\right)-q^{\star}\left(Y\right)\right\Vert _{1}}{\left|\mathcal{Y}\right|}\nonumber \\
 & +D\left(Q_{\boldsymbol{Y}}\left(Y\right)\parallel q^{\star}(Y)\right),\label{eq:sec AEP cond}
\end{alignat}
where $Q_{\boldsymbol{Y}}$ is the type of $\boldsymbol{Y}$. Next
we bound the divergence 
\begin{alignat}{1}
 & D\left(Q_{\boldsymbol{Y}}\left(Y\right)\parallel q^{\star}(Y)\right)=\nonumber \\
 & \sum_{y\in\mathcal{Y}}Q_{\boldsymbol{Y}}\left(y\right)\log\frac{Q_{\boldsymbol{Y}}\left(y\right)}{q^{\star}\left(y\right)}=\nonumber \\
 & \sum_{y\in\mathcal{Y}}\left(\sum_{x\in\mathcal{X}}Q_{\boldsymbol{X},\boldsymbol{Y}}\left(x,y\right)\right)\log\frac{\left(\sum_{x\in\mathcal{X}}Q_{\boldsymbol{X},\boldsymbol{Y}}\left(x,y\right)\right)}{\sum_{x\in\mathcal{X}}q^{\star}\left(x,y\right)}\leq\nonumber \\
 & \sum_{y\in\mathcal{Y},x\in\mathcal{X}}Q_{\boldsymbol{X},\boldsymbol{Y}}\left(x,y\right)\log\frac{Q_{\boldsymbol{X},\boldsymbol{Y}}\left(x,y\right)}{q^{\star}\left(x,y\right)}=\nonumber \\
 & \sum_{y\in\mathcal{Y},x\in\mathcal{X}}Q_{\boldsymbol{X},\boldsymbol{Y}}\left(x,y\right)\log\frac{Q_{\boldsymbol{X}}\left(x\right)P_{\boldsymbol{Y}\mid\boldsymbol{X}}\left(y\mid x\right)}{q^{\star}\left(x\right)p\left(y\mid x\right)}=\nonumber \\
 & D\left(Q_{\boldsymbol{X}}\left(X\right)\parallel q^{\star}\left(X\right)\right)+D\left(P_{\boldsymbol{Y}\mid\boldsymbol{X}}\parallel p\left(Y\mid X\right)\mid Q_{\boldsymbol{X}}\right),
\end{alignat}
where $Q_{\boldsymbol{X},\boldsymbol{Y}}$ is the joint type of $\left(\boldsymbol{X},\boldsymbol{Y}\right)$
and the inequality follows from the log sum inequality. The typicality
of the channel implies that $\forall\gamma_{2}$ there exists $N_{2}$
such that for $N>N_{2}$ 
\begin{equation}
\Pr\left\{ D\left(P_{\boldsymbol{Y}\mid\boldsymbol{X}}\parallel p\left(Y\mid X\right)\mid Q_{\boldsymbol{X}}\right)\geq\gamma_{2}\right\} \leq\eta.\label{eq:Div Py_x py_x typicality}
\end{equation}
Put $\gamma_{3}=\gamma_{0}+\gamma_{2}$ and $\gamma_{4}=\sqrt{2\ln2\gamma_{3}}$.
Since $\forall\epsilon$ there exists $\gamma_{0},\gamma_{2}$ such
that 
\begin{equation}
-\gamma_{4}\log\frac{\gamma_{4}}{\left|\mathcal{Y}\right|}+\gamma_{3}\leq\epsilon,
\end{equation}
from (\ref{eq:Div Qx q star}),(\ref{eq:sec AEP cond}) and (\ref{eq:Div Py_x py_x typicality})
it yields that there exists $N_{y}$ such that for $N>N_{y}$ 
\begin{equation}
\Pr\left\{ \left|-\frac{1}{N}\log q^{\star}(\boldsymbol{Y})-H_{q^{\star}}\left(Y\right)\right|\geq\epsilon\right\} \leq\epsilon/3.
\end{equation}
Using similar arguments, $\forall\epsilon$ there exists $N_{xy}$
such that for $N>N_{xy}$ 
\begin{equation}
\Pr\left\{ \left|-\frac{1}{N}\log q^{\star}(\boldsymbol{X},\boldsymbol{Y})-H_{q^{\star}}\left(X,Y\right)\right|\geq\epsilon\right\} \leq\epsilon/3.
\end{equation}
Thus, $\forall\epsilon$ there exists $N_{\epsilon}>\max\left\{ N_{x},N_{y},N_{xy}\right\} $,
such that 
\begin{equation}
\Pr\left\{ \left(\boldsymbol{X},\boldsymbol{Y}\right)\in A_{\epsilon,q^{\star}\left(X,Y\right)}^{(N)}\right\} \geq1-\epsilon.
\end{equation}
This concludes the proof of first property.

2) If $\tilde{\boldsymbol{X}}$, $\tilde{\boldsymbol{Y}}$ are independent,
then for \emph{$N\geq\max\left(N_{\delta},N_{\epsilon}\right)$}
\begin{alignat}{1}
 & \Pr\left\{ \left(\tilde{\boldsymbol{X}},\tilde{\boldsymbol{Y}}\right)\in A_{\epsilon,q^{\star}\left(X,Y\right)}^{(N)}\right\} =\nonumber \\
 & \sum_{\left(\boldsymbol{x},\boldsymbol{y}\right)\in A_{\epsilon,q^{\star}}^{(N)}}\Pr\left\{ \boldsymbol{x}\mid CC_{II}\right\} \Pr\left\{ \boldsymbol{y}\mid CC_{II}\right\} \leq\nonumber \\
 & \sum_{\left(\boldsymbol{x},\boldsymbol{y}\right)\in A_{\epsilon,q^{\star}}^{(N)}}\left(N+1\right)^{2\left|\mathcal{X}\right|}2^{N2\delta}q^{\star}\left(\boldsymbol{x}\right)q^{\star}\left(\boldsymbol{y}\right)\leq\nonumber \\
 & \left(N+1\right)^{2\left|\mathcal{X}\right|}2^{2\delta N}2^{-N\left(I_{q^{\star}}\left(X;Y\right)-3\epsilon\right)},
\end{alignat}
where the first inequality follows from Lemmas \ref{lem:bounding CCII with CCI}
and \ref{lem:bounding CCII with CCI Y}, and the second follows from
the joint AEP theorem \cite[Theorem 7.6.1]{key-10}, with respect
to the i.i.d. $q^{\star}\left(x,y\right)$. 
\end{IEEEproof}
Hence, Theorem \ref{thm:Joint-AEP-CCII} established the joint AEP
properties of code construction \emph{II} with respect to the set
$A_{\epsilon,q^{\star}\left(X,Y\right)}^{(N)}$. The joint AEP properties
of code construction \emph{II} are enough to show that for $R_{q}<I_{q^{\star}}\left(X;Y\right)$,
the average error probability is tending towards zero with the block
length. This can be shown by applying the channel coding theorem \cite[Theorem 7.7.1]{key-10}
and the joint typicality decoder ($A_{\epsilon,q^{\star}\left(X,Y\right)}^{(N)}$)
to code construction \emph{II}.

According to Theorem \ref{thm:Number of NT codewords}, $R_{s}>D(q^{\star}\left(X\right)\parallel p\left(X\right))$
to ensure that each random set of size $2^{NR_{s}}$ includes a codeword
which satisfies the constraints with very high probability. Therefore,
\begin{equation}
R_{M}\triangleq R_{q}<\min\left\{ H_{p}\left(X\right)-D(q^{\star}\left(X\right)\parallel p\left(X\right)),I_{q^{\star}}\left(X;Y\right)\right\} \label{eq:Matched achievable rate}
\end{equation}
is an achievable rate, taking the maximal rate $R=H_{p}\left(X\right)$. 

Since good codes for uniform probability mass function are readily
available, the special case where the large codebook is uniformly
distributed, i.e. $p\left(x\right)=1/\left|\mathcal{X}\right|,\forall x\in\mathcal{X}$,
is of practical importance. It is easy to show that in this case,
the conditional limit probability mass function $q^{\star}\left(x\right)$
achieves $\max_{P\left(x\right)\in E}H_{P\left(X\right)}$, i.e.,
maximize the entropy within $E$. It yields that, 
\begin{equation}
R_{M}<I_{q^{\star}}\left(X;Y\right),\label{eq:RM_uniform}
\end{equation}
since
\begin{equation}
H_{p}\left(X\right)-D(q^{\star}\left(X\right)\parallel p\left(X\right))=H_{q^{\star}}\left(X\right)
\end{equation}
and $H_{q^{\star}}\left(X\right)>I_{q^{\star}}\left(X;Y\right)$.
This is the achievable rate under matched decoding in this special
case. Since the selected codewords maximize the entropy of the input
and not the mutual information, this is only a lower bound on the
capacity subject to the constraints. Nevertheless, for the average
power constraint on AWGN with discrete constellation, the distribution
which maximizes the capacity is approximately the distribution which
maximizes the entropy. This observation can be verified numerically,
using the modified BAA \cite{key-8}. Hence, the achievable rate is
very close to the maximal rate in this special case.

\section{\label{sec:Mismatched-Decoding}Mismatched Decoding}

The previous sections treated code construction and the achievable
rate under matched decoding. Recall that practical schemes, however,
use a mismatched decoder from complexity considerations. This decoder
is only aware of the large codebook and performs optimal decoding
as if all the codewords might have been transmitted. It is suboptimal,
but occasionally more practical since the adaptation to the shaped
subcode might be extremely complex, and the decoder for the large
code is readily available. This section analyzes the mismatched decoder
beyond the naive approach, which showed that the non-typical codewords
within the large codebook require special treatment. We present two
novel lower bounds for the achievable rate using the mismatched decoder. 

Our first lower bound is based upon Gallager's error exponent bounding
technique \cite{key-11}. The key here is to perform maximum likelihood
decoding using the entire codebook, while taking into account that
only a subcode was transmitted and the true distribution of the transmitted
codewords. Based on this analysis, we redefine the average error probability
and present a new error exponent (\ref{eq:Subcode error prob}). It
is important to note that this bound is suitable for any large codebook
ensemble, as long as it is generated according to i.i.d. distribution
and not necessarily a uniform distribution. Using the new error exponent,
we show that (\ref{eq:achievable transmission rate.}) is an achievable
rate for the mismatched decoder. 

Our second lower bound is based upon the joint typicality decoder.
This decoder is not suitable for non-typical codewords since it declares
an error when the input or the output is not typical. Thus, we define
a modified joint typicality (MJT) decoder, which decodes solely based
on joint typicality between the input and the output with respect
to the large codebook distribution. The MJT achievable rate bound
is presented in (\ref{eq:achievable rate MTP}). It might seems surprising
that decoding according to a wrong distribution might work and indeed
this bound is correct in special cases only (Lemma \ref{lem:First error event}).
One of them is uniform distribution of the large codebook and additive
channels. Thus, the AWGN channel can be treated using this approach.
More examples are shown in the sequel.

\subsection{\label{sec:Gallager-Exponent}Mismatched Decoding: Gallager Exponent}

As already claimed, while the matched decoder achieves higher transmission
rates, the mismatched decoder is more practical. In this section we
use the Gallager bounding technique \cite{key-11} to obtain a lower
bound on the maximal achievable rate for the shaped subcode and the
mismatched decoder. 

The mismatched decoder in this section is the maximum likelihood decoder,
which uses the entire codebook in the decoding process. The average
error probability, however, is redefined (Theorem \ref{thm:Subcode error})
to take into account the true distribution of the transmitted codewords.
Thus, a new error exponent is presented. Using this error exponent,
Theorem \ref{thm:Achievable rate} shows an achievable rate for the
mismatched decoder.

It is important to note that this bound is suitable for any large
codebook ensemble, as long as it is generated according to i.i.d.
distribution $p\left(x\right)$ and not necessarily the uniform distribution. 
\begin{thm}
\label{thm:Subcode error}Let $p\left(y\mid x\right)$ be a memoryless
channel probability mass function. For a given number of codewords
$M=2^{NR}\geq2$ of block length $N$. \textup{\emph{Let the large
code }}ensemble\textup{\emph{ be generated according to i.i.d. $p\left(x\right)$.
Let $E$ be a }}closed convex set\textup{\emph{, }}such that $p\left(x\right)\notin E$.
Let \textup{$q^{\star}\left(x\right)$} be the probability mass function
which achieves $\min_{P\left(x\right)\in E}D\left(P\left(X\right)\parallel p\left(X\right)\right)$
and $R_{s}>D\left(q^{\star}\left(X\right)\parallel p\left(X\right)\right)$.
Define a corresponding ensemble of subcodes, generated by code construction
II with respect to $E$. Suppose that an arbitrary subcode codeword
$m$, $1\leq m\leq M_{q}$, enters the encoder and that maximum-likelihood
decoding using the entire code is employed. Then $\forall\delta$
there exists $N_{\delta}$ such that for $N\geq N_{\delta}$, the
average probability of decoding error over this ensemble of subcodes,
for any choice of $\rho$, $0\leq\rho\leq1$, is bounded by
\begin{align}
 & \overline{P}_{e}\left(E,p\right)\leq\nonumber \\
 & \left(M-1\right)^{\rho}\left(N+1\right)^{\left|\mathcal{X}\right|}2^{N\delta}\left(\sum_{y\in\mathcal{Y}}\left(\sum_{x\in\mathcal{X}}q^{\star}\left(x\right)p\left(y\mid x\right)^{\frac{1}{1+\rho}}\right)\right.\nonumber \\
 & \cdot\left.\left(\sum_{x}p\left(x\right)p\left(y\mid x\right)^{\frac{1}{1+\rho}}\right)^{\rho}\right)^{N}.\label{eq:Subcode error prob}
\end{align}
\end{thm}
\begin{IEEEproof}
Let $\overline{P}_{e,m}\left(E,p\right)$ designate the average error
probability of codeword $m$, when using code construction \emph{II}
to create the subcode ensemble and maximum-likelihood decoding using
the entire code. Let 
\begin{equation}
q\left(\boldsymbol{x}_{m}\right)\triangleq\Pr\left\{ \boldsymbol{x}_{m}\mid CC_{II}\right\} 
\end{equation}
denote the probability of codeword $m$ within the subcode ensemble.
Let $\Pr\left\{ \textrm{error}\mid m,\boldsymbol{x}_{m},\boldsymbol{y},p\left(\boldsymbol{x}\right)\right\} $
be the probability of error conditioned on codeword $m$ entering
the encoder, the particular codeword $\boldsymbol{x}_{m}$, the received
sequence $\boldsymbol{y}$ and $p\left(\boldsymbol{x}\right)$. Then,
the average decoding error probability for the codeword $m$, when
maximum-likelihood decoding using the entire code is employed, is
\begin{align}
 & \overline{P}_{e,m}\left(E,p\right)=\nonumber \\
 & \sum_{\boldsymbol{y}}\sum_{\boldsymbol{x}_{m}}q\left(\boldsymbol{x}_{m}\right)p\left(\boldsymbol{y}\mid\boldsymbol{x}_{m}\right)\Pr\left\{ \textrm{error}\mid m,\boldsymbol{x}_{m},\boldsymbol{y},p\left(\boldsymbol{x}\right)\right\} .\label{eq:avg Pem}
\end{align}
Following the derivation of the Gallager error exponent (proof of
\cite[Theorem 5.6.1]{key-12}), the probability of error given $m$,
$\boldsymbol{x}_{m}$ and $\boldsymbol{y}$ is
\begin{align}
 & \Pr\left\{ \textrm{error}\mid m,\boldsymbol{x}_{m},\boldsymbol{y},p\left(\boldsymbol{x}\right)\right\} \leq\nonumber \\
 & \left(M-1\right)^{\rho}\left(\sum_{\boldsymbol{x}}p\left(\boldsymbol{x}\right)\frac{p\left(\boldsymbol{y}\mid\boldsymbol{x}\right)^{\frac{1}{1+\rho}}}{p\left(\boldsymbol{y}\mid\boldsymbol{x}_{m}\right)^{\frac{1}{1+\rho}}}\right)^{\rho}.\label{eq:Perror givem m xm y}
\end{align}
By substituting (\ref{eq:Perror givem m xm y}) into (\ref{eq:avg Pem})
and recognizing that $\boldsymbol{x}_{m}$ is a dummy variable we
obtain
\begin{align}
 & \overline{P}_{e,m}\left(E,p\right)\leq\left(M-1\right)^{\rho}\cdot\nonumber \\
 & \sum_{\boldsymbol{y}}\left(\sum_{\boldsymbol{x}}q\left(\boldsymbol{x}\right)p\left(\boldsymbol{y}\mid\boldsymbol{x}\right)^{\frac{1}{1+\rho}}\right)\left(\sum_{\boldsymbol{x}}p\left(\boldsymbol{x}\right)p\left(\boldsymbol{y}\mid\boldsymbol{x}\right)^{\frac{1}{1+\rho}}\right)^{\rho}.\label{eq:avg Pem bound}
\end{align}
Using Lemma \ref{lem:bounding CCII with CCI}, $\forall\delta$ there
exists $N_{\delta}$ such that for $N\geq N_{\delta}$, 
\begin{eqnarray}
\Pr\left\{ \boldsymbol{x}\mid CC_{II}\right\}  & \leq & \left(N+1\right)^{\left|\mathcal{X}\right|}2^{N\delta}q^{\star}\left(\boldsymbol{x}\right)\nonumber \\
 & = & \left(N+1\right)^{\left|\mathcal{X}\right|}2^{N\delta}\prod_{i=0}^{N-1}q^{\star}\left(x_{i}\right).
\end{eqnarray}
This yields that, 
\begin{alignat}{1}
 & \overline{P}_{e,m}\left(E,p\right)\leq\left(M-1\right)^{\rho}\left(N+1\right)^{\left|\mathcal{X}\right|}2^{N\delta}\cdot\nonumber \\
 & \sum_{\boldsymbol{y}}\left(\sum_{\boldsymbol{x}}q^{\star}\left(\boldsymbol{x}\right)p\left(\boldsymbol{y}\mid\boldsymbol{x}\right)^{\frac{1}{1+\rho}}\right)\left(\sum_{\boldsymbol{x}}p\left(\boldsymbol{x}\right)p\left(\boldsymbol{y}\mid\boldsymbol{x}\right)^{\frac{1}{1+\rho}}\right)^{\rho}=\nonumber \\
 & \left(M-1\right)^{\rho}\left(N+1\right)^{\left|\mathcal{X}\right|}2^{N\delta}\cdot\nonumber \\
 & \left(\sum_{y}\left(\sum_{x}q^{\star}\left(x\right)p\left(y\mid x\right)^{\frac{1}{1+\rho}}\right)\left(\sum_{x}p\left(x\right)p\left(y\mid x\right)^{\frac{1}{1+\rho}}\right)^{\rho}\right)^{N},
\end{alignat}
where the equality follows since all the probability mass functions
are independent and identically distributed. Each subcode codeword
$m$ is bounded by (\ref{eq:avg Pem bound}), which does not depend
on $m$. Thus, the average probability of decoding error is
\begin{align}
 & \overline{P}_{e}\left(E,p\right)=\sum_{m=1}^{M_{q}}\Pr\left(m\right)\overline{P}_{e,m}\left(E,p\right)\leq\nonumber \\
 & \left(M-1\right)^{\rho}\left(N+1\right)^{\left|\mathcal{X}\right|}2^{N\delta}\cdot\nonumber \\
 & \left(\sum_{y}\left(\sum_{x}q^{\star}\left(x\right)p\left(y\mid x\right)^{\frac{1}{1+\rho}}\right)\left(\sum_{x}p\left(x\right)p\left(y\mid x\right)^{\frac{1}{1+\rho}}\right)^{\rho}\right)^{N},
\end{align}
and we obtain the bound.
\end{IEEEproof}
Thus, the last theorem showed the modified error exponent for the
non-typical codewords. Note that the main difference in the derivation
lays in (\ref{eq:avg Pem}), since the averaging is over the actual
distribution of the codeword $\boldsymbol{x}_{m}$ ($\Pr\left\{ \boldsymbol{x}_{m}\mid CC_{II}\right\} $)
and not the typical codewords in the large codebook. 

Since the number of codewords $M=2^{NR}$ exponentially depends on
the large code rate $R$, $\forall\delta$ with $N\geq N_{\delta}$,
the bound can be expressed as 
\begin{eqnarray*}
\overline{P}_{e}\left(E,p\right) & \leq\left(N+1\right)^{\left|\mathcal{X}\right|} & 2{}^{-N\left(E_{0}\left(\rho,q^{\star},p\right)-\rho R-\delta\right)}
\end{eqnarray*}
where 
\begin{eqnarray}
E_{0}\left(\rho,p_{1},p_{2}\right) & \triangleq & -\log\left(\sum_{y}\left(\sum_{x}p_{1}\left(x\right)p\left(y\mid x\right)^{\frac{1}{1+\rho}}\right)\right.\nonumber \\
 &  & \cdot\left.\left(\sum_{x}p_{2}\left(x\right)p\left(y\mid x\right)^{\frac{1}{1+\rho}}\right)^{\rho}\right).\label{eq:E0qp}
\end{eqnarray}
Once the error exponent is obtained, the achievable rate can be deducted.
The following theorem shows the large code rate $R$, for which the
average error probability is tending towards zero. This rate is later
translated into an achievable transmission rate $R_{q}$, using $R=R_{q}+R_{s}$
(by code construction \emph{II}). 
\begin{thm}
\label{thm:Achievable rate}Suppose that for every $\delta$ there
exists $N\geq N_{\delta}$ such that for every $0\leq\rho\leq1$ and
$E_{0}\left(0,q^{\star},p\right)=0$,
\begin{eqnarray*}
\overline{P}_{e}\left(E,p\right) & \leq\left(N+1\right)^{\left|\mathcal{X}\right|} & 2{}^{-N\left(E_{0}\left(\rho,q^{\star},p\right)-\rho R-\delta\right)}.
\end{eqnarray*}
 Let 
\begin{eqnarray}
I_{q^{\star},p}\left(X,Y\right) & = & \lim_{\rho\rightarrow0}\frac{E_{0}\left(\rho,q^{\star},p\right)}{\rho},\label{eq:MIqp}
\end{eqnarray}
then for every $R<I_{q^{\star},p}\left(X,Y\right)$ and $N\rightarrow\infty$,
the average error probability $\overline{P}_{e}\left(E,p\right)\rightarrow0$. \end{thm}
\begin{IEEEproof}
According to the limit definition, for every $\epsilon>0$ there exists
$\rho>0$ such that 
\begin{equation}
\frac{E_{0}\left(\rho,q^{\star},p\right)}{\rho}>I_{q^{\star},p}\left(X,Y\right)-\epsilon.
\end{equation}
Taking $\epsilon=I_{q^{\star},p}\left(X,Y\right)-R$, this yields
\begin{equation}
E_{0}\left(\rho,q^{\star},p\right)-\rho R=\gamma>0.
\end{equation}
Thus, for every $\epsilon$ the exists $\delta$ and the corresponding
$N_{\delta}$ such that $\gamma>\delta$ and for $N\geq N_{\delta}$
\begin{eqnarray}
0 & \leq & \lim_{N\rightarrow\infty}\overline{P}_{e}\left(E,p\right)\nonumber \\
 & \leq & \lim_{N\rightarrow\infty}\left(N+1\right)^{\left|\mathcal{X}\right|}2{}^{-N\left(E_{0}\left(\rho,q^{\star},p\right)-\rho R-\delta\right)}\nonumber \\
 & = & 0,
\end{eqnarray}
and therefore 
\[
\lim_{N\rightarrow\infty}\overline{P}_{e}\left(E,p\right)=0.
\]

\end{IEEEproof}
Theorem \ref{thm:Achievable rate} showed that when $R<I_{q^{\star},p}\left(X,Y\right)$,
the average error probability $\overline{P}_{e}\left(E,p\right)$
tends towards zero. Furthermore, according to Theorem \ref{thm:Number of NT codewords},
$R_{q}<R-D\left(q^{\star}\left(X\right)\parallel p\left(X\right)\right)$.
Hence, the average error probability $\overline{P}_{e}\left(E,p\right)\rightarrow0$,
for the shaped code if 
\begin{equation}
R_{G}\triangleq R_{q}<I_{q^{\star},p}\left(X,Y\right)-D\left(q^{\star}\left(X\right)\parallel p\left(X\right)\right).
\end{equation}
This is the achievable rate under the mismatched decoding using the
Gallager exponent bounding technique. The closed form of $I_{q^{\star},p}\left(X,Y\right)$
is given at Appendix (\ref{eq:Iqp}). This yield that, 
\begin{align}
R_{G} & <\min\left\{ H_{p}\left(X\right)-D(q^{\star}\left(X\right)\parallel p\left(X\right)),\right.\label{eq:achievable transmission rate.}\\
 & \left.I_{q^{\star}}\left(X;Y\right)+D\left(q^{\star}\left(Y\right)\parallel p\left(Y\right)\right)-D(q^{\star}\left(X\right)\parallel p\left(X\right))\right\} .\nonumber 
\end{align}
is an achievable transmission rate using the Gallager bounding technique.
This relies on the average error probability and the maximal number
of codewords using code construction \emph{II,} which was discussed
in Section \ref{sub:Number-of-codewords}. We use the following lemma
to verify that the mismatched decoder performs is worse than the matched.
\begin{lem}
Let $p\left(y\mid x\right)$ be memoryless channel and $p_{1}\left(x\right)$,
$p_{2}\left(x\right)$ be input distributions. Let $p_{1}\left(y\right)$,
$p_{2}\left(y\right)$ be the corresponding channel outputs, such
that $p_{i}\left(y\right)=\sum_{x\in\mathcal{X}}p_{i}\left(x\right)p\left(y\mid x\right)$.
Then,
\begin{equation}
D\left(p_{1}\left(Y\right)\parallel p_{2}\left(Y\right)\right)\leq D\left(p_{1}\left(X\right)\parallel p_{2}\left(X\right)\right).
\end{equation}
\end{lem}
\begin{IEEEproof}
According to the definition of divergence
\begin{align}
 & D\left(p_{1}\left(Y\right)\parallel p_{2}\left(Y\right)\right)=\nonumber \\
 & \sum_{y\in\mathcal{Y}}p_{1}\left(y\right)\log\frac{p_{1}\left(y\right)}{p_{2}\left(y\right)}=\nonumber \\
 & \sum_{y\in\mathcal{Y}}\left(\sum_{x\in\mathcal{X}}p_{1}\left(x\right)p\left(y\mid x\right)\right)\log\frac{\left(\sum_{x\in\mathcal{X}}p_{1}\left(x\right)p\left(y\mid x\right)\right)}{\left(\sum_{x\in\mathcal{X}}p_{2}\left(x\right)p\left(y\mid x\right)\right)}\leq\nonumber \\
 & \sum_{y\in\mathcal{Y},x\in\mathcal{X}}p_{1}\left(x\right)p\left(y\mid x\right)\log\frac{p_{1}\left(x\right)p\left(y\mid x\right)}{p_{2}\left(x\right)p\left(y\mid x\right)}=\nonumber \\
 & \sum_{x\in\mathcal{X}}p_{1}\left(x\right)\log\frac{p_{1}\left(x\right)}{p_{2}\left(x\right)}=\nonumber \\
 & D\left(p_{1}\left(X\right)\parallel p_{2}\left(X\right)\right),
\end{align}
where the inequality follows from the log-sum inequality.
\end{IEEEproof}
Using $p_{1}\left(x\right)=q^{\star}\left(x\right)$ and $p_{2}\left(x\right)=p\left(x\right)$,
we conclude that 
\begin{equation}
D\left(q^{\star}\left(Y\right)\parallel p\left(Y\right)\right)\leq D\left(q^{\star}\left(X\right)\parallel p\left(X\right)\right);
\end{equation}
therefore $R_{G}<I_{q^{\star}}\left(X;Y\right)$ as expected.

\subsection{\label{sec:Joint-Typicality-Decoder}Modified Joint Typicality Decoder}

Our second lower bound for mismatched decoding is based on the joint
typicality decoder analysis. In the original configuration, however,
this decoder is not suitable for non-typical codewords in $\mathcal{C}$,
since it declares an error when the input or the output is not typical
according to $p\left(x,y\right)$. Thus, we define a modified joint
typicality decoder. The MJT decoder is aware only of the large codebook
$\mathcal{C}$, and decodes solely based on joint typicality between
the input and the output with respect to the large codebook probability
mass function $p\left(x,y\right)$. It might seems surprising that
decoding according to a wrong probability mass function might work
and indeed this bound is correct in special cases only. Lemma \ref{lem:First error event},
provides a sufficient condition on $p\left(x\right)$ and the channel,
such that the MJT decoder is applicable. We restrict our analysis
to this particular case. We bound the average error probability of
the shaped subcode $\mathcal{C}_{s}\in\mathcal{C}$, created by code
construction \emph{II}, using the MJT decoder to obtain an achievable
rate. 

Suppose that a codeword $\boldsymbol{X}\left(1\right)\in\mathcal{C}_{s}$
with index $m=1$ was sent through the channel and a noisy sequence
$\boldsymbol{Y}$ was received. The original joint typicality decoder
operates as follows; it examines all the codewords in the large codebook
and establishes which codewords are jointly typical with the received
sequence $\boldsymbol{Y}$. Since, $\boldsymbol{X}\left(1\right)$
and $\boldsymbol{Y}$ are typical with respect to $q^{\star}\left(x\right)$
and $q^{\star}\left(y\right)$, respectively; they are not typical
with respect to $p\left(x\right)$ and $p\left(y\right)$. Therefore,
the joint typicality decoder declares an error for each transmitted
codeword. For this reason, the typicality of $\boldsymbol{X}\left(1\right)$
and $\boldsymbol{Y}$ with respect to $p\left(x\right)$, $p\left(y\right)$
is disregarded.

Let us define the modified joint typicality set and the modified joint
typicality decoder.
\begin{defn}
\label{M_pxy}The modified jointly typical set $M_{\epsilon,p\left(x,y\right)}^{(N)}$
with respect to the probability mass function $p\left(x,y\right)$
is the set of sequences $(\boldsymbol{x},\boldsymbol{y})\in\mathcal{X}^{N}\times\mathcal{Y}^{N}$
with empirical entropies $\epsilon$-close to the true entropy $H_{p}\left(X,Y\right)$:
\end{defn}
\begin{eqnarray}
M_{\epsilon,p\left(x,y\right)}^{(N)} & = & \left\{ (\boldsymbol{x},\boldsymbol{y})\in\mathcal{X}^{N}\times\mathcal{Y}^{N}:\right.\nonumber \\
 &  & \left.\left|-\frac{1}{N}\log p(\boldsymbol{x},\boldsymbol{y})-H_{p}\left(X,Y\right)\right|<\epsilon\right\} .\label{eq:MJTP set}
\end{eqnarray}
The modified joint typicality decoder operates as follows; first it
examines all the codewords in the large codebook and establishes which
codewords are MJT with the received sequence $\boldsymbol{Y}$, i.e.,
belong to the set $M_{\epsilon,p\left(x,y\right)}^{(N)}$. An error
occurs, either when the sent codeword $\boldsymbol{X}\left(1\right)$
is not MJT with the received sequence $\boldsymbol{Y}$ or when more
that one codeword is MJT with $\boldsymbol{Y}$. 

Let us define the event that a codeword $\boldsymbol{X}\left(i\right)\in\mathcal{C}$
is MJT with the output as $E_{i}$, $i=\left\{ 1,...,2^{nR}\right\} $.
The event $E_{1}^{c}$ (the complete to $E_{1}$) is the first error
event, i.e., the sent codeword is not MJT with the received sequence.
Thus, the probability of error is therefore the union of these events.
Let us now define an event $E_{Y_{q}}$, where the received sequence
$\boldsymbol{Y}$ is typical with respect to the probability mass
function $q^{\star}\left(y\right)$, i.e., $\boldsymbol{Y}\in A_{\epsilon,q^{\star}\left(y\right)}^{(N)}$
and $E_{Y_{q}}^{c}$ as the complete to $E_{Y_{q}}$. Using the AEP
property for code construction \emph{II} 
\begin{equation}
\Pr\left\{ E_{Y_{q}}^{c}\mid m=1\right\} <\frac{\epsilon}{2},
\end{equation}
for sufficiently large $N$. Hence, the probability of error is bounded

\begin{align}
 & \Pr\left\{ \textrm{error}|m=1\right\} =\nonumber \\
 & \Pr\left\{ E_{1}^{c}\cup E_{2}\cup\cdots\cup E_{2^{nR}}|m=1\right\} =\nonumber \\
 & \Pr\left\{ \left(E_{1}^{c}\cup E_{2}\cup\cdots\cup E_{2^{nR}}\right)\cap E_{Y_{q}}|m=1\right\} \Pr\left\{ E_{Y_{q}}|m=1\right\} +\nonumber \\
 & \Pr\left\{ \left(E_{1}^{c}\cup E_{2}\cup\cdots\cup E_{2^{nR}}\right)\cap E_{Y_{q}}^{c}|m=1\right\} \Pr\left\{ E_{Y_{q}}^{c}|m=1\right\} \leq\nonumber \\
 & \Pr\left\{ \left(E_{1}^{c}\cup E_{2}\cup\cdots\cup E_{2^{nR}}\right)\cap E_{Y_{q}}|m=1\right\} +\frac{\epsilon}{2}\leq\nonumber \\
 & \Pr\left\{ E_{1}^{c}|m=1\right\} +\sum_{i=2}^{2^{NR}}\Pr\left\{ E_{i}\cap E_{Y_{q}}|m=1\right\} +\frac{\epsilon}{2},\label{eq:error prob bound}
\end{align}
where the last inequality follows from the union bound and since $\Pr\left\{ E_{1}^{c}\cap E_{Y_{q}}|m=1\right\} \leq\Pr\left\{ E_{1}^{c}|m=1\right\} $.

First, we analyze $\Pr\left\{ E_{1}^{c}|m=1\right\} $, which is equal
to $\Pr\left\{ \left(\boldsymbol{X}\left(1\right),\boldsymbol{Y}\right)\notin M_{\epsilon,p\left(x,y\right)}^{(N)}\right\} $.
The following lemma introduces a sufficient condition such that this
probability is sufficiently small. In the general case, however, the
decoder will not choose the right codeword, since the probability
of joint typicality according to a wrong probability mass function
($p\left(x,y\right)$) is extremely low.
\begin{lem}
\label{lem:First error event}For sufficiently large $N$
\begin{equation}
\Pr\left\{ \left(\boldsymbol{X}\left(1\right),\boldsymbol{Y}\right)\notin M_{\epsilon,p\left(x,y\right)}^{(N)}\right\} <\epsilon
\end{equation}
if 
\begin{equation}
\log p\left(x\right)-H_{p}\left(Y\mid X=x\right)=c,\;\forall x\in\mathcal{X},\label{eq:Entropy equality}
\end{equation}
where $c$ is a constant.\end{lem}
\begin{IEEEproof}
Since $p\left(\boldsymbol{X}\left(1\right),\boldsymbol{Y}\right)=\prod_{i=0}^{N-1}p\left(X_{i}\left(1\right),Y_{i}\right)$,
it yields that
\begin{align}
 & -\frac{1}{N}\log p\left(\boldsymbol{X}\left(1\right),\boldsymbol{Y}\right)=\nonumber \\
 & -\frac{1}{N}\sum_{i=0}^{N-1}\log p\left(X_{i}\left(1\right),Y_{i}\right)=\nonumber \\
 & -\sum_{x\in\mathcal{X},y\in\mathcal{Y}}Q_{\boldsymbol{X}\left(1\right),\boldsymbol{Y}}\left(x,y\right)\log p(x,y)=\nonumber \\
 & -\sum_{x\in\mathcal{X},y\in\mathcal{Y}}Q_{\boldsymbol{X}\left(1\right)}\left(x\right)P_{\boldsymbol{Y}\mid\boldsymbol{X}\left(1\right)}\left(x,y\right)\log p(x,y)=\nonumber \\
 & -\sum_{x\in\mathcal{X}}Q_{\boldsymbol{X}\left(1\right)}\left(x\right)\left[\log p(x)-H_{P_{\boldsymbol{Y}\mid\boldsymbol{X}\left(1\right)}}\left(Y\mid X=x\right)\right.\nonumber \\
 & \left.-D\left(P_{\boldsymbol{Y}\mid\boldsymbol{X}\left(1\right)}\left(Y\mid X=x\right)\parallel p\left(Y\mid X=x\right)\right)\right],
\end{align}
where $Q_{\boldsymbol{X}\left(1\right),\boldsymbol{Y}}$ is the type
of $\left(\boldsymbol{X}\left(1\right),\boldsymbol{Y}\right)$ and
$P_{\boldsymbol{Y}\mid\boldsymbol{X}\left(1\right)}$ is the type
of the channel, i.e., of $\boldsymbol{Y}$ given $\boldsymbol{X}\left(1\right)$.
Hence, 
\begin{alignat}{1}
 & \left|-\frac{1}{N}\log p\left(\boldsymbol{X}\left(1\right),\boldsymbol{Y}\right)-H_{p}\left(X,Y\right)\right|=\nonumber \\
 & \left|\sum_{x\in\mathcal{X}}\left(p\left(x\right)-Q_{\boldsymbol{X}\left(1\right)}\left(x\right)\right)\left[\log p\left(x\right)-H_{p}\left(Y\mid X=x\right)\right]\right.+\nonumber \\
 & \sum_{x\in\mathcal{X}}Q_{\boldsymbol{X}\left(1\right)}\left(x\right)\left[H_{P_{\boldsymbol{Y}\mid\boldsymbol{X}\left(1\right)}}\left(Y\mid X=x\right)-H_{p}\left(Y\mid X=x\right)\right.+\nonumber \\
 & \left.\left.D\left(P_{\boldsymbol{Y}\mid\boldsymbol{X}\left(1\right)}\left(Y\mid X=x\right)\parallel p\left(Y\mid X=x\right)\right)\right]\right|\leq\nonumber \\
 & \sum_{x\in\mathcal{X}}Q_{\boldsymbol{X}\left(1\right)}\left(x\right)\left[\left|H_{P_{\boldsymbol{Y}\mid\boldsymbol{X}\left(1\right)}}\left(Y\mid X=x\right)-H_{p}\left(Y\mid X=x\right)\right|\right.+\nonumber \\
 & \left.D\left(P_{\boldsymbol{Y}\mid\boldsymbol{X}\left(1\right)}\left(Y\mid X=x\right)\parallel p\left(Y\mid X=x\right)\right)\right],
\end{alignat}
where the last derivation is correct since $\log p\left(x\right)-H_{p}\left(Y\mid X=x\right)=c$
for every $x\in\mathcal{X}$. Using the typicality of the channel
and the same arguments as in Theorem \ref{thm:Joint-AEP-CCII}, for
every $\epsilon$ there exists $N_{\epsilon}$ such that for $N>N_{\epsilon}$
\begin{alignat}{1}
 & \left|H_{P_{\boldsymbol{Y}\mid\boldsymbol{X}\left(1\right)}}\left(Y\mid X=x\right)-H_{p}\left(Y\mid X=x\right)\right|+\nonumber \\
 & D\left(P_{\boldsymbol{Y}\mid\boldsymbol{X}\left(1\right)}\left(Y\mid X=x\right)\parallel p\left(Y\mid X=x\right)\right)\leq\epsilon,
\end{alignat}
for every $x\in\mathcal{X}$ with probability larger than $1-\epsilon$.
We conclude that for $N>N_{\epsilon}$
\begin{equation}
\Pr\left\{ \left|-\frac{1}{N}\log p\left(\boldsymbol{X}\left(1\right),\boldsymbol{Y}\right)-H_{p}\left(X,Y\right)\right|>\epsilon\right\} \leq\epsilon,
\end{equation}
i.e., $\Pr\left\{ \left(\boldsymbol{X}\left(1\right),\boldsymbol{Y}\right)\notin M_{\epsilon,p\left(x,y\right)}^{(N)}\right\} <\epsilon$.
\end{IEEEproof}
A special case where condition (\ref{eq:Entropy equality}) is satisfied
is uniform probability mass function $p\left(x\right)=1/\left|\mathcal{X}\right|,\;\forall x\in\mathcal{X}$
and a channel which satisfies $H_{p}\left(Y\mid X=x\right)=c,\;\forall x$.
Binary symmetric channel (BSC) and additive noise channels $Y=X+N$,
are examples for which this property is satisfied. Hence, for these
channels and uniformly distributed large codebook 
\begin{equation}
\Pr\left\{ E_{1}^{c}|m=1\right\} <\frac{\epsilon}{2},\label{eq:first error event}
\end{equation}
for sufficiently large $N$. 

Next we bound the probability $\Pr\left\{ E_{i}\cap E_{Y_{q}}|m=1\right\} $,
$i=2,\ldots,2^{nR}$. The event $E_{i}\cap E_{Y_{q}}|m=1$ is equivalent
to $\left(\boldsymbol{X}\left(i\right),\boldsymbol{Y}\right)\in M_{\epsilon,p\left(x,y\right)}^{(N)}\cap A_{\epsilon,q^{\star}\left(y\right)}^{(N)}$.
Since by the code generation process, $\boldsymbol{X}\left(1\right)$
and $\boldsymbol{X}\left(i\right)$ are independent for $i\neq1$,
so are $\boldsymbol{Y}$ and $\boldsymbol{X}\left(i\right)$ for $i\neq1$.
Since $\boldsymbol{X}\left(i\right)$ was generated according to i.i.d.
$p\left(x\right)$ and $\boldsymbol{Y}$ was generated from code construction
\emph{II} 
\begin{equation}
\Pr\left\{ \boldsymbol{X}\left(i\right),\boldsymbol{Y}\right\} =p\left(\boldsymbol{X}\left(i\right)\right)\Pr\left\{ \boldsymbol{Y}\mid CC_{II}\right\} .
\end{equation}
Thus, using Lemma \ref{lem:bounding CCII with CCI Y} 
\begin{alignat}{1}
 & \Pr\left\{ E_{i}\cap E_{Y_{q}}|m=1\right\} =\nonumber \\
 & \sum_{\left(\boldsymbol{x},\boldsymbol{y}\right)\in M_{\epsilon,p\left(x,y\right)}^{(N)}\cap A_{\epsilon,q^{\star}\left(y\right)}^{(N)}}\Pr\left\{ \boldsymbol{x},\boldsymbol{y}\right\} =\nonumber \\
 & \sum_{\left(\boldsymbol{x},\boldsymbol{y}\right)\in M_{\epsilon,p\left(x,y\right)}^{(N)}\cap A_{\epsilon,q^{\star}\left(y\right)}^{(N)}}p\left(\boldsymbol{x}\right)\Pr\left\{ \boldsymbol{y}\mid CC_{II}\right\} \leq\nonumber \\
 & \left(N+1\right)^{\left|\mathcal{X}\right|}2^{N\delta}\sum_{\left(\boldsymbol{x},\boldsymbol{y}\right)\in M_{\epsilon,p\left(x,y\right)}^{(N)}\cap A_{\epsilon,q^{\star}\left(y\right)}^{(N)}}p\left(\boldsymbol{x}\right)q^{\star}\left(\boldsymbol{y}\right)=\nonumber \\
 & \left(N+1\right)^{\left|\mathcal{X}\right|}2^{N\delta}\Pr\left\{ E_{i}\cap E_{Y_{q}}|m=1,CC_{I}\right\} .
\end{alignat}
The event $E_{i}\cap E_{Y_{q}}|m=1,CC_{I}$ is easier to analyze.
It is equivalent to the event $P_{\boldsymbol{X}\left(i\right),\boldsymbol{Y}}\in\Lambda\cap\mathcal{P}_{N}\left(X,Y\right)$,
where $\left(\boldsymbol{X}\left(i\right),\boldsymbol{Y}\right)$
are drawn according to i.i.d. $p\left(x\right)q^{\star}\left(y\right)$,
$P_{\boldsymbol{X}\left(i\right),\boldsymbol{Y}}$ is the type of
$\left(\boldsymbol{X}\left(i\right),\boldsymbol{Y}\right)$ and 
\begin{eqnarray}
\Lambda & = & \left\{ P:\right.\nonumber \\
 &  & \left|\sum_{x\in\mathcal{X},y\in\mathcal{Y}}P\left(x,y\right)\log p(x,y)-H_{p}\left(X,Y\right)\right|\leq\epsilon,\nonumber \\
 &  & \left.\left|\sum_{x\in\mathcal{X},y\in\mathcal{Y}}P\left(x,y\right)\log q^{\star}(y)-H_{q}\left(Y\right)\right|\leq\epsilon\right\} .\label{eq:pi constraints}
\end{eqnarray}
According to Sanov's theorem \cite[Theorem 11.4.1]{key-11}, the probability
that $\left(\boldsymbol{X}\left(i\right),\boldsymbol{Y}\right)$ has
type $P_{\boldsymbol{X}\left(i\right),\boldsymbol{Y}}\in\Lambda\cap\mathcal{P}_{N}\left(X,Y\right)$
is bounded by 
\begin{equation}
\Pr\left\{ E_{i}\cap E_{Y_{q}}|m=1,CC_{I}\right\} \leq2^{-ND\left(P^{\star}\parallel p\left(X\right)q^{\star}\left(Y\right)\right)}\label{eq:second error event}
\end{equation}
where
\begin{equation}
P^{\star}\left(x,y\right)=\arg\min_{P\left(x,y\right)\in\Lambda}D\left(P\parallel p\left(X\right)q^{\star}\left(Y\right)\right).\label{eq:Pstar-MJT}
\end{equation}
To obtain the minimal divergence in $\Lambda$, we construct the following
Lagrangian and its derivative
\begin{align}
J\left(P\right) & =D\left(P\Vert p\left(X\right)q^{\star}(Y)\right)\\
 & +\;\lambda_{0}\sum P\left(x,y\right)\nonumber \\
 & +\;\lambda_{1}\sum P\left(x,y\right)\log q^{\star}\left(y\right)\nonumber \\
 & +\;\lambda_{2}\sum P\left(x,y\right)\log p\left(x,y\right),\nonumber \\
\frac{dJ\left(P\left(x,y\right)\right)}{dP\left(x,y\right)} & =\log\frac{P\left(x,y\right)}{p\left(x\right)q^{\star}(y)}+\lambda_{0}\\
 & +\lambda_{1}\log q^{\star}\left(y\right)+\lambda_{2}\log p\left(x,y\right).\nonumber 
\end{align}
The solution is given in a parametric form 
\begin{align}
 & P^{\star}\left(x,y\right)=\label{eq:Opt P}\\
 & p\left(x\right)q^{\star}(y)\exp\left(\lambda_{0}+\lambda_{1}\log q^{\star}\left(y\right)+\lambda_{2}\log p\left(x,y\right)\right),\nonumber 
\end{align}
where the Lagrange multipliers $\lambda_{0},\lambda_{1},\lambda_{2}$
should satisfy the constraints in (\ref{eq:pi constraints}). 

Finally, by substituting (\ref{eq:first error event}) and (\ref{eq:second error event})
into (\ref{eq:error prob bound}), the error probability can be further
bounded
\begin{equation}
\Pr\left\{ \textrm{error}|m=1\right\} \leq\epsilon+2^{-N\left(D\left(P^{\star}\parallel p\left(X\right)q^{\star}\left(Y\right)\right)-R-\delta\right)}.
\end{equation}
Hence, for 
\begin{eqnarray}
R & = & R_{q}+R_{s}\nonumber \\
 & < & D\left(P^{\star}\left(X,Y\right)||p\left(X\right)q^{\star}\left(Y\right)\right),
\end{eqnarray}
the error probability tend towards zero for $N\rightarrow\infty$.
According to Theorem \ref{thm:Number of NT codewords}, $R_{s}>D(q^{\star}\left(X\right)\parallel p\left(X\right))$
to ensure that each random set of size $2^{NR_{s}}$ includes a codeword
which satisfies the constraints with very high probability. It yields
that for $N\rightarrow\infty$, the minimum between
\begin{alignat}{1}
 & R_{MJT}\triangleq R_{q}<\nonumber \\
 & D\left(P^{\star}\left(X,Y\right)||p\left(X\right)q^{\star}\left(Y\right)\right)-D(q^{\star}\left(X\right)\parallel p\left(X\right))\label{eq:achievable rate MTP}
\end{alignat}
and the maximal amount of codewords (\ref{eq:maximal number of code words})
is an achievable rate using this bounding technique.

\section{\label{sec:Special-cases}Special cases}

In the previous sections we discussed code construction \emph{II},
as a method to obtain shaped codebooks. We have shown the maximal
amount of the available codewords and achievable rates for the matched
and the mismatched decoders. In this section we analyze several special
cases. The first is the case where the large codebook is generated
according to an i.i.d. uniform distribution and the channel satisfies
the condition in Lemma \ref{lem:First error event}. In this case,
all the bounds are applicable. The more general case, where the large
codebook is non-uniform, is later discusses in context of several
special cases: the BNSC and AWGN with Gaussian large codebook. The
details are given next.

\subsection{Uniform large codebook}

The special case where the large codebook is uniformly distributed
on the support is of practical importance. Furthermore, although the
Gallager bound can be obtained for any i.i.d. $p\left(x\right)$,
the MJT bound can be obtained only when condition (\ref{eq:Entropy equality})
is satisfied. In order to compare all the bounds, the following examples
focus on a special case of (\ref{eq:Entropy equality}), where $p\left(x\right)=1/\left|\mathcal{X}\right|$
and the channel satisfies 
\begin{equation}
H_{p}\left(Y\mid X=x\right)=c,\;\forall x\in\mathcal{X}\label{eq:Symmetry cond}
\end{equation}
for any $c$. In particular, we analyze the AWGN and BSC channels. 

Our first observation is regarding the achievable rate for the matched
decoder. It is easy to show that for uniform $p\left(x\right)$, the
conditional limit probability mass function $q^{\star}\left(x\right)$
does not maximize the mutual information, but the entropy $H_{P}\left(X\right)$
within the set $E$, i.e.,
\begin{equation}
q^{\star}\left(x\right)=\arg\max_{P\left(x\right)\in E}H_{P}\left(X\right).
\end{equation}
Thus, code construction \emph{II} can not achieve the maximal rate
even for the matched decoder, since generally different probability
mass functions maximize the mutual information and the entropy subject
to the constraints. Nevertheless, for the average power constraint
over AWGN with discrete input constellation, the probability mass
function which maximizes the mutual information is approximately the
probability mass function which maximizes the entropy. Hence, the
achievable rate is very close to the maximal rate. This can be verified
using the modified BAA \cite{key-8}. 

For the mismatched decoder, the Gallager bound takes the form of
\begin{equation}
R_{G}<I_{q^{\star}}\left(X;Y\right)+D\left(q^{\star}\left(Y\right)\parallel p\left(Y\right)\right)+H_{q^{\star}}\left(X\right)-\log\left|\mathcal{X}\right|,\label{eq:Gallager rate uniform}
\end{equation}
since
\begin{equation}
D\left(q^{\star}\left(X\right)\parallel p\left(X\right)\right)=\log\left|\mathcal{X}\right|-H_{q^{\star}}\left(X\right).\label{eq:uniform Rs}
\end{equation}
The parametric form of the probability mass function $P^{\star}\left(x,y\right)$
of the MJT bound in (\ref{eq:Opt P}) is also simplified into
\begin{eqnarray}
P^{\star}\left(x,y\right) & = & \mathcal{K}q^{\star}(y)^{\tilde{\lambda}_{1}}p\left(y\mid x\right)^{\lambda_{2}},\label{eq:Opt P uniform}
\end{eqnarray}
where $\tilde{\lambda}_{1}=\lambda_{1}+1$, $\lambda_{2}$ and $\mathcal{K}$
(represents all the constants) satisfy the constraints in (\ref{eq:pi constraints}).
There is no closed form solution in the general case and the Lagrange
multipliers should be obtained numerically. Once these parameters
are found, the MJT bound takes the form of
\begin{equation}
R_{MJT}<D\left(P^{\star}\left(X,Y\right)||p\left(X\right)q^{\star}\left(Y\right)\right)+H_{q^{\star}}\left(X\right)-\log\left|\mathcal{X}\right|.\label{eq:MJT rate uniform}
\end{equation}

The maximal amount of codewords is not a limitation in this case,
since the entropy of the large codebook is $\log\left|\mathcal{X}\right|$.
Thus taking the maximal $R=\log\left|\mathcal{X}\right|$, we can
obtain the typical set of $q^{\star}\left(x\right)$ (up to the $\delta$),
since $R_{q}<H_{q^{\star}}\left(X\right)$. For more details recall
Section \ref{sub:Number-of-codewords}.

\subsubsection{AWGN }

The first example that we analyze is the memoryless AWGN channel with
uniformly distributed large codebook, which is a special case of (\ref{eq:Symmetry cond}).
Let $\mathcal{X}$ be a discrete constellation with cardinality $\left|\mathcal{X}\right|$.\textbf{
}Let $\boldsymbol{X}=\left(X_{0},\cdots,X_{N-1}\right)$ be a random
vector of length $N$, representing the transmitted symbols, where
$X_{i}\in\mathcal{X}$. We allow the amplitude levels of the constellation
points to change to model an amplifier at the transmitter. This is
done by multiplying the constellation by a real and positive scaling
parameter $\alpha>0$. Let $\boldsymbol{Z}=\left(Z_{0},\cdots,Z_{N-1}\right)$
be an i.i.d. Gaussian noise vector, where $Z_{i}\sim\mathcal{N}(0,1)$
are i.i.d. Gaussian noise samples, normalized to unit variance. Let
$\boldsymbol{Y}=\left(Y_{0},\cdots,Y_{N-1}\right)$ be the random
vector at the receiver, such that
\begin{equation}
Y_{i}=\alpha X_{i}+Z_{i}.
\end{equation}
The transmission is subject to an average power constraint, where
the maximum average power allowed is $\beta_{0}$. Since the scaling
of the constellation is a degree of freedom, for each scaling $\alpha$
the power constraint can be formulated as
\begin{equation}
E\left(\alpha\right)=\left\{ P:\sum_{x\in\mathcal{X}}P\left(x\right)\varphi_{0}\left(x,\alpha\right)\leq\beta_{0}\right\} ,
\end{equation}
using $\varphi_{0}\left(x,\alpha\right)=\alpha^{2}\left|x\right|^{2}$.
The corresponding decision rule for choosing codewords is therefore
\begin{equation}
\frac{1}{N}\sum_{i=0}^{N-1}\left|\alpha X_{i}\right|^{2}\leq\beta_{0},
\end{equation}
which is equivalent to 
\begin{equation}
\frac{1}{N}\sum_{i=0}^{N-1}\left|X_{i}\right|^{2}\leq\beta_{0}/\alpha^{2}.\label{eq:AVG_decission_rule}
\end{equation}
For example, let the constellation be $\mathcal{X}=\left\{ -3,-1,1,3\right\} $
(4-PAM), and the power constraint be $\beta_{0}=5$. For each scaling
$\alpha$, $E\left(\alpha\right)$ is a set of probability mass functions
which satisfy the constraint. In particular, lets consider $\alpha=1,1.1,1.3$
and the corresponding sets $E\left(1\right)$, $E\left(1.1\right)$
and $E\left(1.3\right)$. 
\begin{figure}
\begin{centering}
\includegraphics[width=8cm,height=8cm,keepaspectratio]{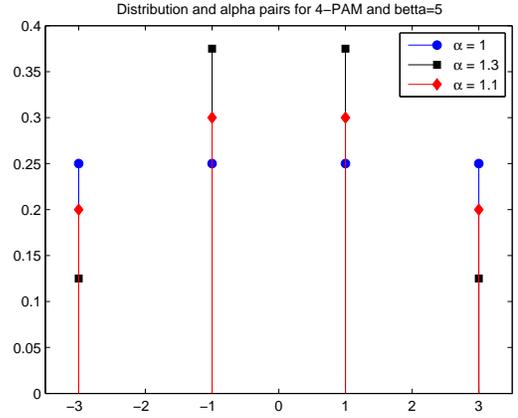}
\par\end{centering}

\caption{\label{fig:Probability-mass-functions}Probability mass functions
corresponding to $\alpha=1,1.1,1.3$ and $\beta=5$.}
\end{figure}
The probability mass functions 
\begin{eqnarray}
p_{1} & = & \left(\frac{1}{4},\frac{1}{4},\frac{1}{4},\frac{1}{4}\right),
\end{eqnarray}
\begin{equation}
p_{1.1}=\left(\frac{2}{10},\frac{3}{10},\frac{3}{10},\frac{2}{10}\right),
\end{equation}
\begin{equation}
p_{1.3}=\left(\frac{1}{8},\frac{3}{8},\frac{3}{8},\frac{1}{8}\right),
\end{equation}
in Fig. \ref{fig:Probability-mass-functions} satisfy the constraints
in $E\left(1\right)$, $E\left(1.1\right)$ and $E\left(1.3\right)$,
respectively. It can be seen that larger scaling is required as the
distribution becomes more {}``Gaussian''. Thus, there is a tradeoff
between the scaling parameter and how much the distribution is {}``Gaussian''
or non-uniform.

Thus, we have proved that (\ref{eq:RM_uniform}), (\ref{eq:Gallager rate uniform})
and (\ref{eq:MJT rate uniform}) are achievable rates for the matched
and mismatched decoders, corresponding to large code distribution
$p\left(x\right)$ and the constraints sets $E$. Since $\alpha$
is a degree of freedom, we still need to establish the optimal scaling
parameter $\alpha^{\star}$, which attains the maximum of each bound.
This also dictates the optimal decision rule to choose the codewords
in the shaped code construction process. Hence, the achievable rate
using the matched decoder is
\begin{equation}
R_{Mu}=\max_{\alpha}I_{q_{\alpha}^{\star}\left(x\right)}\left(X;Y\right).\label{eq:CMU rate}
\end{equation}
It is also known that the Maxwell Boltzmann distribution achieves
the maximal entropy subject an average power constraint \cite{key-10},
therefore the conditional limit probability mass function is 
\begin{equation}
q_{\alpha}^{\star}\left(x\right)=\frac{\exp\left(-t\left|x\right|^{2}\right)}{\sum_{x^{\prime}\in\mathcal{X}}\exp\left(-t\left|x^{\prime}\right|^{2}\right)}.\label{eq:MB dist}
\end{equation}
The parameter $t$ is chosen such that the constraint is satisfied
with equality, i.e.,
\begin{equation}
\sum_{x\in\mathcal{X}}q_{\alpha}^{\star}\left(x\right)\varphi\left(x,\alpha\right)=\beta_{0},\label{eq:AWGN const}
\end{equation}
which follows from the concavity of the entropy.

The maximal rates for the mismatched Gallager and MJT bounds are also
calculated with respect to the optimal $\alpha$ for each bound, such
that
\begin{alignat}{1}
 & R_{Gu}<\label{eq:RGU rate}\\
 & \max_{\alpha}I_{q_{\alpha}^{\star}}\left(X;Y\right)+D\left(q_{\alpha}^{\star}\left(Y\right)\parallel p\left(Y\right)\right)+H_{q_{\alpha}^{\star}}\left(X\right)-\log\left|\mathcal{X}\right|\nonumber 
\end{alignat}
and 
\begin{alignat}{1}
 & R_{MJTu}<\label{eq:CMJT}\\
 & \max_{\alpha}D\left(P_{\alpha}^{\star}\left(X,Y\right)||p\left(X\right)q_{\alpha}^{\star}\left(Y\right)\right)+H_{q_{\alpha}^{\star}}\left(X\right)-\log\left|\mathcal{X}\right|,\nonumber 
\end{alignat}
using (\ref{eq:MB dist}).

\begin{figure}
\begin{centering}
\includegraphics[width=8cm,height=8cm,keepaspectratio]{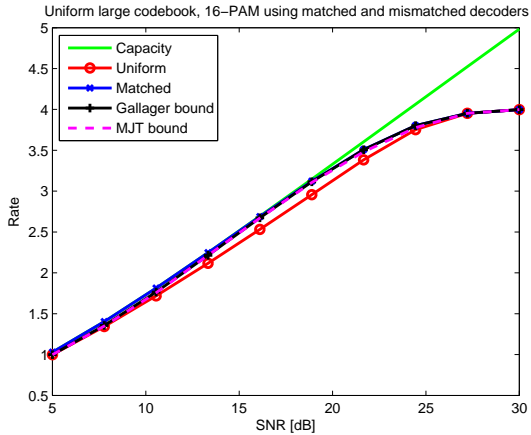}
\par\end{centering}

\caption{\label{fig:AWGN16PAM}Achievable rates 16-PAM, AWGN }
\end{figure}
 
\begin{figure}
\begin{centering}
\includegraphics[width=8cm,height=8cm,keepaspectratio]{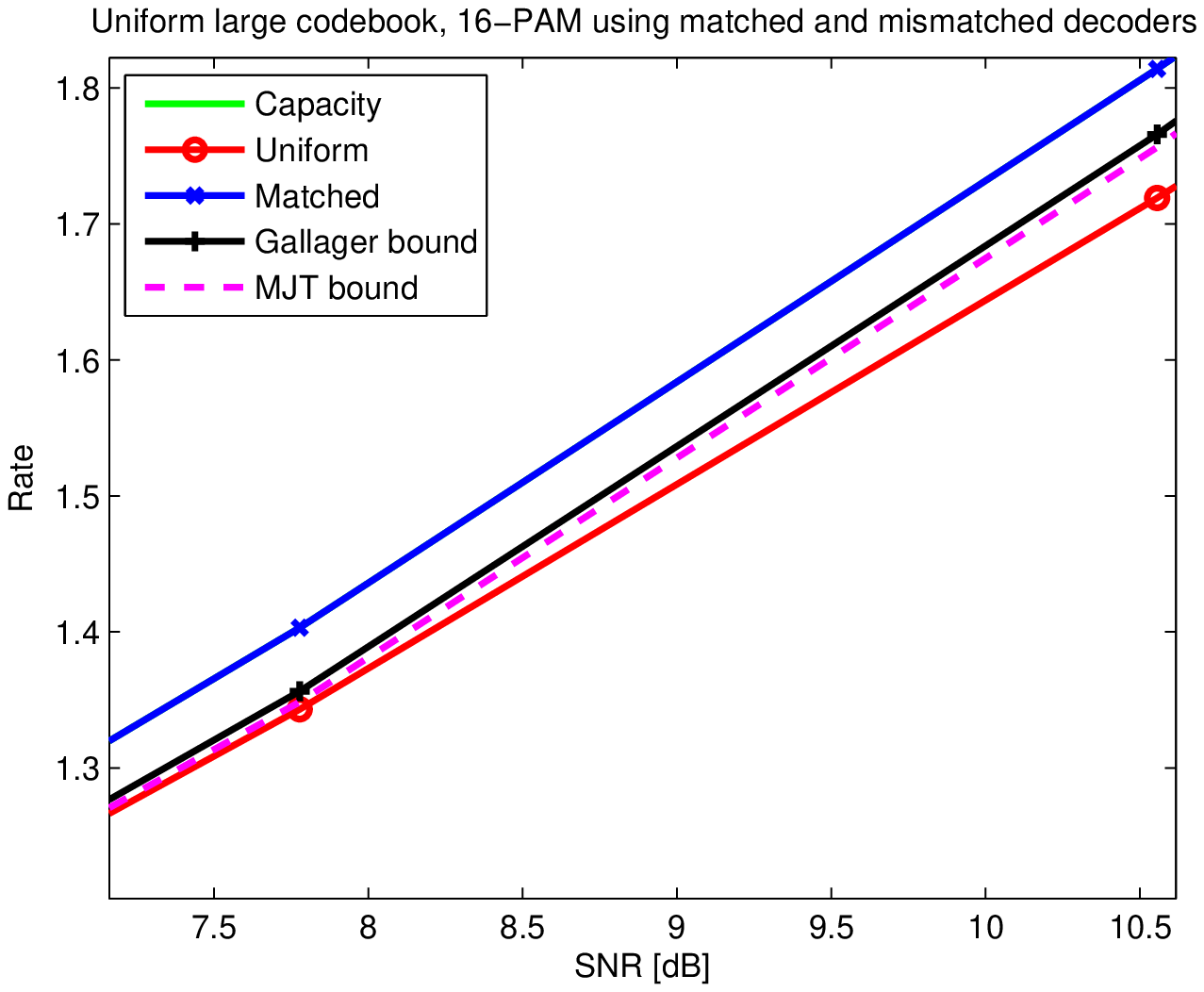}
\par\end{centering}

\caption{\label{fig:AWGN16PAM - Low SNR}Achievable rates 16-PAM, AWGN, low
SNR.}
\end{figure}
 
\begin{figure}
\begin{centering}
\includegraphics[width=8cm,height=8cm,keepaspectratio]{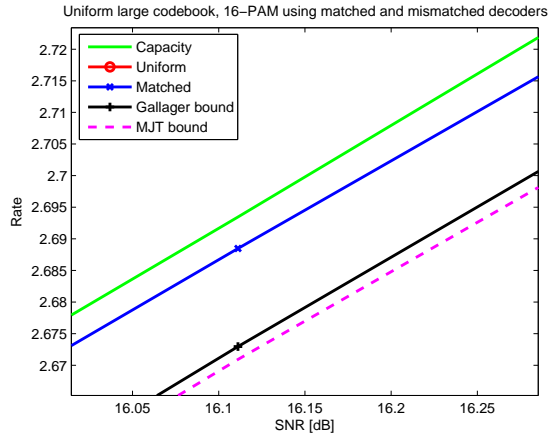}
\par\end{centering}

\caption{\label{fig:AWGN16PAM - High SNR}Achievable rates 16-PAM, AWGN, high
SNR.}
\end{figure}
 Figs. \ref{fig:AWGN16PAM}-\ref{fig:AWGN16PAM - High SNR} present
the maximal achievable rates using the matched decoder and our two
lower bounds for the mismatched decoder versus SNR. The scaling parameter
$\alpha$ is optimized for each bound at each SNR point. We keep the
noise variance constant ($Z_{i}\sim\mathcal{N}(0,1)$) and change
the transmission power $\beta_{0}$ to control the SNR. The large
codebook is uniformly distributed over 16-PAM. These rates are compared
to the capacity of the power constrained continuous AWGN channel,
which is denoted by {}``Capacity''. They are further compared to
the uniformly distributed 16-PAM constellation, which is denoted by
{}``Uniform''. It can be seen that the matched decoder shaping curve
is very close to the capacity curve up to the rate of 3 bps. The gap
to capacity in high SNR (around 3 bps) is approximately 0.1 dB, 0.2
dB for the matched and the mismatched decoders, respectively. In low
SNR the gap is approximately 0 dB, 0.3 dB for the matched and the
mismatched decoders, respectively. Thus, the main conclusion from
the graphs is that for PAM over AWGN there is no significant loss
due to mismatched decoding, and that shaping gain can be attained.
This conclusion actually justifies the works of Forney and others
(unknowingly), who used mismatched decoding.

\subsubsection{BSC }

Our second example is the binary symmetric channel, in which the input
$\mathcal{X}=\left\{ 0,1\right\} $ is inverted with probability $\gamma$.
The mutual information of this channel is
\begin{equation}
I_{P}\left(X;Y\right)=H_{P}\left(Y\right)-H\left(\gamma\right),
\end{equation}
where $P$ is the input probability mass function and $H\left(\gamma\right)=-\left(\gamma\log\gamma+\left(1-\gamma\right)\log\left(1-\gamma\right)\right)$.
\begin{figure}
\centering{}\includegraphics[width=8cm,height=8cm,keepaspectratio]{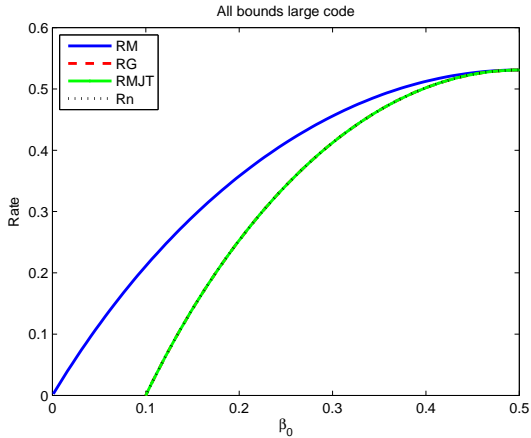}\caption{\label{fig:BSC}BSC: matched and the mismatched decoder versus $\beta_{0}$,
for $\gamma=0.1$ and uniform large codebook.}
\end{figure}
 This channel is another special case of (\ref{eq:Symmetry cond}),
for $P=0.5$ (uniform large codebook). The hamming power constraint
\begin{equation}
\frac{1}{N}\sum_{i=0}^{N-1}X_{i}\leq\beta_{0},
\end{equation}
is very common for this channel. Thus, let $E$ be the constrained
probability mass functions set
\begin{equation}
E=\left\{ P:\sum_{x\in\left\{ 0,1\right\} }P\left(x\right)x\leq\beta_{0}\right\} ,\label{eq:hamming power const}
\end{equation}
where $\beta_{0}\leq\frac{1}{2}$ is the maximal number of '1's allowed,
and let $q^{\star}\left(x\right)$ be the conditional limit probability
mass function which attains minimum $D\left(P\left(X\right)\parallel p\left(X\right)\right)$
in $E$. Using the convexity of the divergence in $P\left(x\right)$,
it implies that the constraint is satisfied with an equality. Thus,
$q^{\star}\left(x=1\right)=\beta_{0}$. The achievable rate for the
matched decoder is therefore $I_{q^{\star}}\left(X;Y\right)=H_{q^{\star}}\left(Y\right)-H\left(\gamma\right)$.

Next we obtain the achievable rates $R_{G}$ and $R_{MJT}$ for this
channel, using the Gallager and the MJT bounding techniques. The Gallager
bound is simplified to 
\begin{eqnarray}
R_{G} & < & H_{q^{\star}}\left(X\right)-H\left(\gamma\right).
\end{eqnarray}
 The rate $R_{MJT}$ may be obtained numerically, using the parametric
form in (\ref{eq:Opt P uniform}). Alternatively, let us guess that
the Lagrange multipliers are $\tilde{\lambda}_{1}=1,\lambda_{2}=1$,
with $\mathcal{K}=1$. Subsequently, 
\begin{equation}
P^{\star}\left(x,y\right)=q^{\star}(y)p\left(y|x\right)
\end{equation}
satisfies the constraints in $\Lambda$. This can be easily verified
by 
\begin{align}
 & \sum_{x\in\mathcal{X},y\in\mathcal{Y}}P^{\star}\left(x,y\right)\log p(x,y)=\nonumber \\
 & \sum_{x\in\mathcal{X},y\in\mathcal{Y}}q^{\star}(y)p\left(y|x\right)\log\frac{1}{2}p\left(y|x\right)=\nonumber \\
 & H\left(\gamma\right)-1=H_{p}\left(X,Y\right)
\end{align}
and
\begin{eqnarray}
P^{\star}\left(y\right) & = & \sum_{x\in\mathcal{X}}q^{\star}(y)p\left(y|x\right)\nonumber \\
 & = & q^{\star}(y)\sum_{x\in\mathcal{X}}p\left(y|x\right)=q^{\star}(y),
\end{eqnarray}
which assures that the second constraint is satisfied. Hence, the
rate
\begin{eqnarray}
R_{MJT} & = & D\left(P^{\star}\parallel p\left(X\right)q\left(Y\right)\right)+H_{q^{\star}}\left(X\right)-\log\left|\mathcal{X}\right|\nonumber \\
 & = & H_{q^{\star}}\left(X\right)-H\left(\gamma\right)
\end{eqnarray}
is achievable and it coincides with the Gallager bound. 

The large code has error probability tending towards zero if $R\leq I_{p}\left(X;Y\right)$
and therefore 
\begin{eqnarray}
R_{n} & < & I_{p}\left(X;Y\right)+H_{q^{\star}}\left(X\right)-\log\left|\mathcal{X}\right|\nonumber \\
 & = & H_{q^{\star}}\left(X\right)-H\left(\gamma\right),\label{eq:naiive approach}
\end{eqnarray}
using (\ref{eq:uniform Rs}), $I_{p}\left(X,Y\right)=1-H\left(\gamma\right)$
and $\log\left|\mathcal{X}\right|=1$. It follows that the three bounds
(Gallager, MJT and the naive approach) coincide in this particular
case. It is known that the BSC capacity is attained for uniform input
probability mass function, and that linear codes achieve the BSC capacity.
Such codes have the same distance spectrum for all the codewords and
therefore the same error probability. In particular, the error probability
of the non-typical codewords equals to the error probability of the
typical. This example provides some insight to why the naive approach
is achievable in this case (as shown using Gallager and MJT). The
rate loss is 
\begin{equation}
R_{loss}=H_{q^{\star}}\left(Y\right)-H_{q^{\star}}\left(X\right),
\end{equation}
comparing to the matched decoder. Results are illustrated in Fig.
\ref{fig:BSC}, using $\gamma=0.1$.

\subsubsection{BNSC }

The third example is the binary non-symmetric channel. This channel
is an example where condition (\ref{eq:Symmetry cond}) is not satisfied.
Thus, for the practical case of uniform large codebook, only the Gallager
bound can be obtained. 

Let the input be $\mathcal{X}=\left\{ 0,1\right\} $. The input is
complemented with probability $\gamma_{0}$ if $x=0$ and $\gamma_{1}$
if $x=1$. The achievable rate of this channel is
\begin{equation}
I_{P}\left(X;Y\right)=H_{P}\left(Y\right)-PH\left(\gamma_{0}\right)-\left(1-P\right)H\left(\gamma_{1}\right),
\end{equation}
where $P\triangleq P\left(x=0\right)$ is the input probability mass.
\begin{figure}
\centering{}\includegraphics[width=8cm,height=8cm,keepaspectratio]{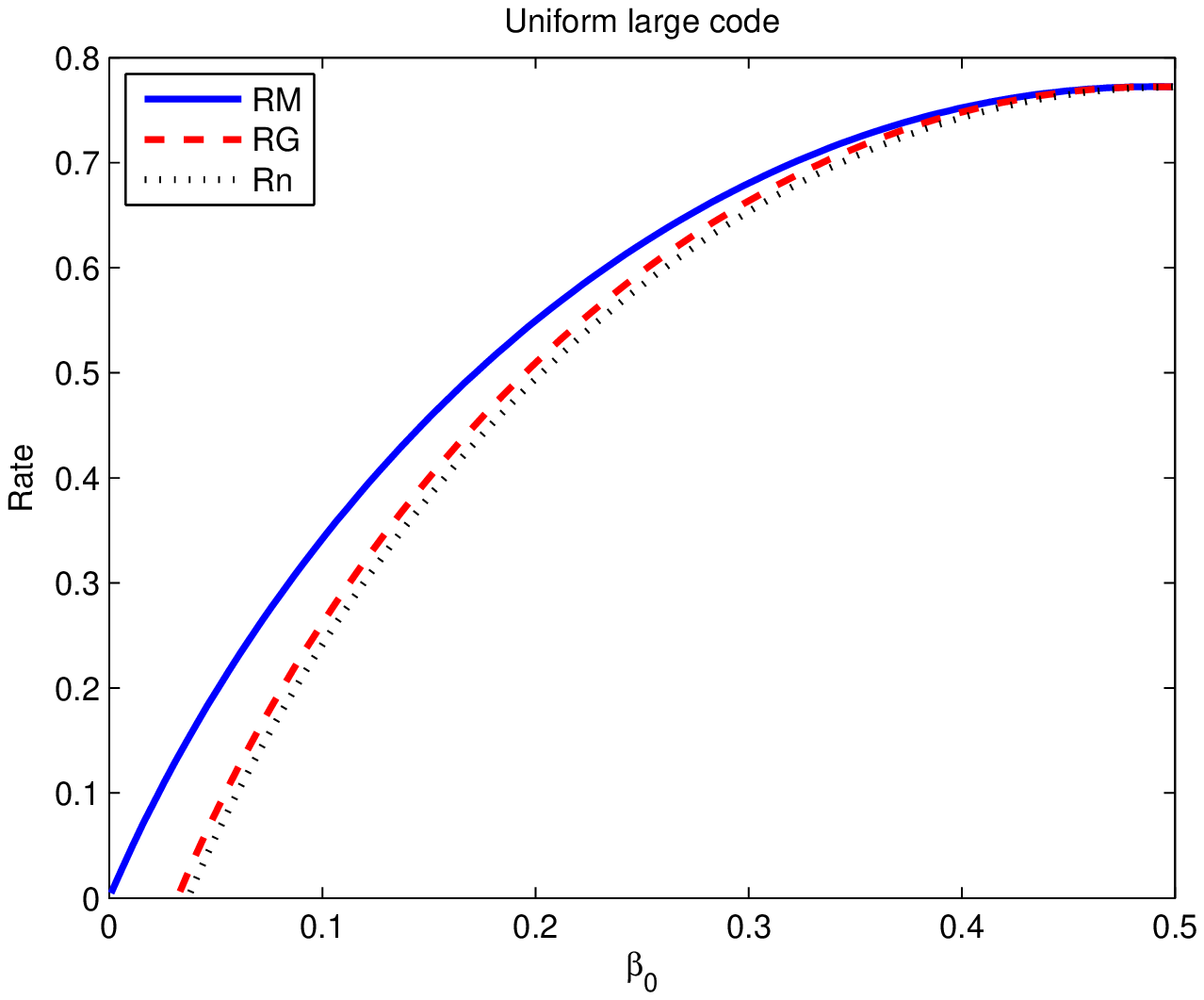}\caption{\label{fig:BNSC_Uniform-1}BNSC: matched and the mismatched decoder
versus $q^{\star}$, for $\gamma_{0}=0.025$ , $\gamma_{1}=0.05$
and $p=0.5$.}
\end{figure}
Let the hamming power constraint be the same as for the BSC channel
(\ref{eq:hamming power const}). Fig. \ref{fig:BNSC_Uniform-1} shows
the achievable rates for the matched and mismatched decoders for $\gamma_{0}=0.025$
and $\gamma_{1}=0.05$. The matched decoder outperforms the mismatched
as expected. At $\beta_{0}=0.5$, the typical codewords are transmitted
since the large codebook is uniform, thus the two decoders are the
same. The naive approach describes the performance of the typical
codewords. In this case, the mismatched performs better than the naive
approach, thus the non-typical codewords have lower average probability
of error than the typical codewords.

\subsection{Non-uniform large codebook}

The non-uniform large codebook is the more general case. Although
it is not common to use non-uniform distributions in practical system,
we present several toy examples for the sake of completeness. The
first example is the BNSC which satisfies the general condition (\ref{eq:Entropy equality})
(required for the MJT bound to be valid). Otherwise, we can not compare
all the bounds. The second example is generating a Gaussian codebook
with power $\beta_{0}$ from a larger Gaussian codebook with average
power $B_{0}$, where $\beta_{0}\ll B_{0}$.

\subsubsection{BNSC }

In this section we consider the binary non-symmetric channel, such
that the general condition (\ref{eq:Entropy equality}) is satisfied.
In order to be able to demonstrate all our bounds (i.e., satisfy (\ref{eq:Entropy equality}))
we have to choose a large codebook which is not uniformly distributed.
In particular, the large codebook is according to $p\left(x\right)$
which solves
\begin{equation}
\left\{ \begin{array}{c}
\log p\left(x=0\right)-H\left(\gamma_{0}\right)=\log p\left(x=1\right)-H\left(\gamma_{1}\right),\\
p\left(x=0\right)+p\left(x=1\right)=1,
\end{array}\right.
\end{equation}
i.e., 
\begin{equation}
p\left(x=0\right)=\frac{\exp\left(H\left(\gamma_{0}\right)-H\left(\gamma_{1}\right)\right)}{\exp\left(H\left(\gamma_{0}\right)-H\left(\gamma_{1}\right)\right)+1}.\label{eq:pBNSC}
\end{equation}
The hamming power constraint is used once again, limiting the average
number of ones in a codeword to be lower than in the large codebook,
i.e. $\beta_{0}\leq p\left(x=1\right)$. 
\begin{figure}
\begin{centering}
\includegraphics[width=8cm,height=8cm,keepaspectratio]{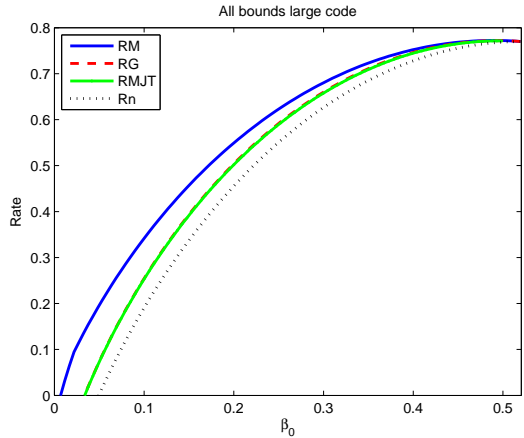}
\par\end{centering}

\caption{\label{fig:BNSC}BNSC: matched and the mismatched decoder versus $\beta_{0}$,
for $\gamma_{0}=0.025$ , $\gamma_{1}=0.05$ and $p=0.47$.}
\end{figure}
Fig. \ref{fig:BNSC} shows the achievable rates for $\gamma_{0}=0.025$
and $\gamma_{1}=0.05$, using the matched and the mismatched decoders.
The probability mass function of the large codebook (according to
(\ref{eq:pBNSC})) is $p=0.47$. In this example, the Gallager and
the MJT are also above the naive approach. Furthermore, the Gallager
and the MJT bounds are very close to each other. All bounds coincide
at $q^{\star}=0.47$, since the typical codewords are transmitted.

\subsubsection{Gaussian large code }

In the following example we consider the situation where the large
codebook has $2^{NR}$ codewords with average power $B_{0}$, according
to Gaussian distribution. Whereas, the transmitted codewords have
much smaller power $\beta_{0}\ll B_{0}$. It is easy to show that
the conditional distribution in this case is also Gaussian with average
power $\beta_{0}$. Thus, the capacity of the Gaussian channel is
achieved for the matched decoder. 

The mismatched decoder derivation is less trivial. Let us first derive
the Gallager bound using
\begin{align}
 & D\left(q^{\star}\left(X\right)\parallel p\left(X\right)\right)=\nonumber \\
 & \frac{1}{2}\left(\frac{\beta_{0}}{B_{0}}-\log\frac{\beta_{0}}{B_{0}}-1\right),
\end{align}
where $q^{\star}\left(X\right)$ is Gaussian with variance $\beta_{0}$
and zero mean, and $p\left(X\right)$ is Gaussian with variance $B_{0}$
and zero mean. In a similar manner,

\begin{align}
 & D\left(q^{\star}\left(Y\right)\parallel p\left(Y\right)\right)=\nonumber \\
 & \frac{1}{2}\left(\frac{\beta_{0}+\sigma_{Z}^{2}}{B_{0}+\sigma_{Z}^{2}}-\log\frac{\beta_{0}+\sigma_{Z}^{2}}{B_{0}+\sigma_{Z}^{2}}-1\right),
\end{align}
assuming that the channel is AWGN with variance $\sigma_{Z}^{2}$
and zero mean. In the scenario where $\beta_{0}\ll B_{0}$ and $\sigma_{Z}^{2}\ll B_{0}$,
the Gallager bound is deduced to
\begin{eqnarray}
R_{G} & \approx & \frac{1}{2}\log\left(1+\frac{\beta_{0}}{\sigma_{Z}^{2}}\right)+\frac{1}{2}\log\left(\frac{\beta_{0}}{\beta_{0}+\sigma_{Z}^{2}}\right)\nonumber \\
 & = & \frac{1}{2}\log\left(\beta_{0}/\sigma_{Z}^{2}\right).\label{eq:GallagerGaussian}
\end{eqnarray}
This example is related to nested lattice shaping. The nested lattice
scheme consists of fine lattice and coarse lattice. The fine lattice
is the {}``large codebook'' (having infinite number of codewords).
This lattice represents all the cosets. The coarse lattice provides
the shaping region, such that it bounds the coset representatives
(the minimum power codewords in each coset), which construct the shaped
codebook. A practical decoding scheme for nested lattices is lattice
decoding, which is in fact very similar to the mismatched decoder
described in this paper. The large Gaussian codebook can be viewed
as a region in the infinite lattice, which is large enough such that
decoding in that codebook is the same as decoding in infinite lattice
for codewords transmitted from the small codebook. It is interesting
though to compare our bound (\ref{eq:GallagerGaussian}) to the performance
of lattice decoding \cite{key-16}, and indeed the result is identical.

\section{Conclusions}

In this paper, we proposed a method to analyze practical shaping schemes,
in which a shaped subcode is chosen from a larger code (in most cases
uniformly distributed large codebook). The codewords in the shaped
codebook satisfy as set of constraints. We provide a theoretical framework
for such analysis, using random coding and a selection process subject
to constraints. In particular, we have found the achievable rates
of the matched and the mismatched decoders, where the last is unaware
of the selection process at the transmitter. This decoder is suboptimal,
but occasionally more practical since it does not have to repeat the
selection process at the receiver. In general, we show that the large
code performance does not dictate the subcode performance, and that
the non-typical codewords require different analysis. Hence, we obtained
two novel achievable bounds for the mismatched decoding using a modification
on the Gallager and the Joint Typicality bounding techniques. We examine
several special cases in detail. In the special case of BSC, however,
the large code analysis is actually valid and can be justified using
capacity achieving linear codes. For M-PAM over AWGN the loss due
to mismatched decoding is not very significant. We show other examples
where there is a larger difference between the matched and mismatched
decoding.

\appendix{}

To obtain $I_{q^{\star},p}\left(X,Y\right)$ in closed form, we calculate
the derivative of $E_{0}\left(\rho,q^{\star},p\right)$ with respect
to $\rho$. Let us define
\begin{eqnarray}
a\left(\rho,y\right) & = & \sum_{x}q^{\star}\left(x\right)p\left(y\mid x\right)^{\frac{1}{1+\rho}},\label{eq:a}\\
b\left(\rho,y\right) & = & \sum_{x}p\left(x\right)p\left(y\mid x\right)^{\frac{1}{1+\rho}},\label{eq:b}\\
c\left(\rho,y\right) & = & b\left(\rho,y\right)^{\rho}\nonumber \\
 & = & \exp\left(\rho\log b\left(\rho,y\right)\right),\label{eq:c}\\
d\left(\rho\right) & = & \sum_{y}a\left(\rho,y\right)c\left(\rho,y\right).\label{eq:d}
\end{eqnarray}
Substituting (\ref{eq:a})-(\ref{eq:d}) into (\ref{eq:E0qp}) 
\begin{equation}
E_{0}\left(\rho,q,p\right)=-\log d\left(\rho\right).\label{eq:param E0}
\end{equation}
Differentiating with respect to $\rho$, yields 
\begin{align}
 & \frac{\partial E_{0}\left(\rho,q,p\right)}{\partial\rho}=-\frac{\partial}{\partial\rho}\log d\left(\rho\right)\nonumber \\
= & -\frac{\sum_{y}\frac{\partial a\left(\rho,y\right)}{\partial\rho}c\left(\rho,y\right)+\frac{\partial c\left(\rho,y\right)}{\partial\rho}a\left(\rho,y\right)}{d\left(\rho\right)},\label{eq:diff E0qp}
\end{align}
such that 
\begin{equation}
\frac{\partial a\left(\rho,y\right)}{\partial\rho}=\frac{-\sum_{x}q^{\star}\left(x\right)p\left(y\mid x\right)^{\frac{1}{1+\rho}}\log p\left(y\mid x\right)}{\left(1+\rho\right)^{2}}
\end{equation}
\begin{equation}
\frac{\partial b\left(\rho,y\right)}{\partial\rho}=\frac{-\sum_{x}p\left(x\right)p\left(y\mid x\right)^{\frac{1}{1+\rho}}\log p\left(y\mid x\right)}{\left(1+\rho\right)^{2}}
\end{equation}
\begin{equation}
\frac{\partial c\left(\rho,y\right)}{\partial\rho}=c\left(\rho,y\right)\left(\log b\left(\rho,y\right)+\rho\frac{\frac{\partial b\left(\rho,y\right)}{\partial\rho}}{b\left(\rho,y\right)}\right).\label{eq:diff E0qp 2}
\end{equation}
Finally, using these terms at $\rho=0$, yields
\begin{align}
 & \frac{\partial E_{0}\left(\rho,q,p\right)}{\partial\rho}\mid_{\rho=0}=\nonumber \\
 & \sum_{y}\sum_{x}q^{\star}\left(x,y\right)\log p\left(y\mid x\right)\nonumber \\
 & -\sum_{y}q^{\star}\left(y\right)\log p\left(y\right).
\end{align}
Subsequently, using the definitions of entropy (\ref{eq:entropy}),
conditional entropy (\ref{eq:conditional entropy}), mutual information
(\ref{eq:mutual information}) and the Kullback-Leibler information
divergence (\ref{eq:divergence}) we conclude that 
\begin{equation}
I_{q^{\star},p}\left(X,Y\right)=I_{q^{\star}}\left(X,Y\right)+D\left(q^{\star}\left(Y\right)\parallel p\left(Y\right)\right).\label{eq:Iqp}
\end{equation}
This gives $I_{q^{\star},p}\left(X,Y\right)$ in closed form.

\end{document}